\documentclass[runningheads]{llncs}
\usepackage{paralist}
\usepackage{units}
\usepackage{booktabs} % For formal tables
\usepackage{graphicx}
\usepackage{latexsym}
\usepackage{times}
 \usepackage{multirow}
\usepackage{booktabs}
\usepackage{xspace}
\usepackage{cite}
\usepackage{subcaption}
\usepackage[all,pdftex]{xy}
\SelectTips{cm}{}

\newif\ifignore % when set to true, additional text appears containing
\ignorefalse
\newcommand{\auxproof}[1]{
\ifignore\mbox{}\newline
\textbf{BEGIN: AUX-PROOF} \dotfill\newline
{#1}\mbox{}\newline
\textbf{END: AUX-PROOF}\dotfill\newline
\fi}

\renewcommand{\phi}{\varphi}

\newcommand{\R}{\mathbb{R}}
\newcommand{\N}{\mathbb{N}}
\newcommand{\Rpos}{\R_{+}}

\newcommand{\RN}{\R^{N}}
\newcommand{\RM}{\R^{M}}

\newcommand{\M}{\mathcal{M}}

\newcommand{\MR}{\mathcal{R}}

\newcommand{\argmax}{\mathop{\rm arg~max}\limits}

\newcommand{\bv}{\mathbf{v}}
\newcommand{\bw}{\mathbf{w}}
\newcommand{\bu}{\mathbf{u}}
\newcommand{\sem}[1]{\llbracket #1 \rrbracket}

\newcommand{\DiaOp}[1]{\Diamond_{#1}}
\newcommand{\BoxOp}[1]{\Box_{#1}}
\newcommand{\speed}{\mathit{speed}}
\newcommand{\gear}{\mathit{gear}}
\newcommand{\rpm}{\mathit{rpm}}
\newcommand{\AF}{\mathit{AF}}
\newcommand{\AFref}{\mathit{AFref}}
\newcommand{\contrMode}{\mathit{controller\_mode}}
\newcommand{\muSpec}{\mathit{mu}}

\newcommand{\STL}{\textrm{STL}}

\newcommand{\Var}{\mathbf{Var}}

\newcommand{\UntilOp}[1]{\mathbin{\mathcal{U}_{#1}}}

\newcommand{\Rnn}{\R_{\ge 0}}

\newcommand{\Defeq}{:=}
\newcommand{\Robust}[2]{{ \llbracket #1, #2 \rrbracket}}

\newcommand{\Vee}[1]{{{\bigsqcup_{#1}}}}
\newcommand{\Wedge}[1]{{{\bigsqcap_{#1}}}}
\newcommand{\throttle}{\mathit{throttle}}
\newcommand{\brake}{\mathit{brake}}

\newcommand{\rew}{\mathsf{rew}}

\newcommand{\mabUcb}{\texttt{MAB-UCB}\xspace}
\newcommand{\mabEps}{\texttt{MAB-\-$\epsilon$-\-greedy}\xspace}

\newcommand{\fr}{SR\xspace}
\newcommand{\breach}{{\tt Breach}\xspace}
\newcommand{\basicBench}{{\tt Bbench}\xspace}
\newcommand{\scaledBench}{{\tt Sbench}\xspace}

\spnewtheorem{mytheorem}{Theorem}%[section]
{\bfseries}{\itshape} 
\spnewtheorem{mylemma}[mytheorem]{Lemma}{\bfseries}{\itshape}
\spnewtheorem{myproposition}[mytheorem]{Proposition}{\bfseries}{\itshape}
\spnewtheorem{mysublemma}[mytheorem]{Sublemma}{\bfseries}{\itshape}
\spnewtheorem{mycorollary}[mytheorem]{Corollary}{\bfseries}{\itshape}
\spnewtheorem{myfact}[mytheorem]{Fact}{\bfseries}{\itshape}
\spnewtheorem{mynotation}[mytheorem]{Notation}{\bfseries}{\rmfamily}
\spnewtheorem{myremark}[mytheorem]{Remark}{\bfseries}{\rmfamily}
\spnewtheorem{myexample}[mytheorem]{Example}{\bfseries}{\rmfamily}
\spnewtheorem{myassumption}[mytheorem]{Assumption}{\bfseries}{\rmfamily}
\spnewtheorem{mydefinition}[mytheorem]{Definition}{\bfseries}{\rmfamily}
\spnewtheorem{myrequirements}[mytheorem]{Requirements}{\bfseries}{\rmfamily}
\spnewtheorem{myproblem}[mytheorem]{Problem}{\bfseries}{\rmfamily}
\usepackage{algorithm}
\usepackage[noend]{algpseudocode} 
\algnewcommand{\IIf}[1]{\State\algorithmicif\ #1\ \algorithmicthen}
\algnewcommand{\EndIIf}{\unskip\ %\algorithmicend\ \algorithmicif
}

\usepackage{bm}
\usepackage{amsmath}
\usepackage{amssymb}
\usepackage{amsfonts}
\usepackage{mathtools}
\usepackage{stmaryrd}
\usepackage{graphicx}
\usepackage{color}
\usepackage{colortbl}

\usepackage{comment}
\usepackage{proof}
\usepackage{fancybox}

\usepackage{wrapfig}

\usepackage{pgfplotstable} 
\pgfplotsset{compat=1.12}

\usepackage{tabularx}

\usepackage{arydshln}
\usepackage{hyperref}
\hypersetup{
colorlinks = false, % false: boxed links; true: colored links
hidelinks = true,
linkcolor=black, % color of internal links
citecolor=black, % color of links to bibliography
urlcolor=black, % color of external links
filecolor=black
}

\usepackage{enumitem}

\usepackage{multirow}
\usepackage{pifont}

\newcounter{researchquestionCount}
\newcommand{\researchquestion}[1]{\stepcounter{researchquestionCount}\vspace{10pt}\noindent\parbox{0.97\textwidth}{{\bf RQ\arabic{researchquestionCount}} {\it #1}}\vspace{0pt}}

\begin{document}

\title{Multi-Armed Bandits for Boolean Connectives in Hybrid System Falsification (Extended Version)\thanks{This is the extended author version of the manuscript with the same name published in the proceedings of the 31st International Conference on Computer-Aided Verification (CAV 2019). The final version is available at \url{www.springer.com}. The authors are supported by ERATO HASUO Metamathematics for Systems Design Project (No. JPMJER1603), JST.}}

\titlerunning{Multi-Armed Bandits for Boolean Connectives in Hybrid System Falsification}

\author{Zhenya Zhang\inst{1,2}%\orcidID{0000-0002-3854-9846}
\and
Ichiro Hasuo\inst{1,2}%\orcidID{0000-0002-8300-4650}
\and
Paolo Arcaini\inst{1}%\orcidID{0000-0002-6253-4062}
}

\authorrunning{Z. Zhang et al.}

\institute{National Institute of Informatics, Tokyo, Japan\\
\email{\{zhangzy,hasuo,arcaini\}@nii.ac.jp}
\and
 SOKENDAI (The Graduate University for Advanced Studies), Hayama, Japan
}

\maketitle

\begin{abstract}
\emph{Hybrid system falsification} is an actively studied topic, as a scalable quality assurance methodology for real-world cyber-physical systems. In falsification, one employs stochastic hill-climbing optimization to quickly find a counterexample input to a black-box system model. Quantitative \emph{robust semantics} is the technical key that enables use of such optimization. In this paper, we tackle the so-called \emph{scale problem} regarding Boolean connectives that is widely recognized in the community: quantities of different scales (such as speed [km/h] vs.\ rpm, or worse, rph) can mask each other's contribution to robustness. Our solution consists of integration of the \emph{multi-armed bandit} algorithms in hill climbing-guided falsification frameworks, with a technical novelty of a new reward notion that we call \emph{hill-climbing gain}. Our experiments show our approach's robustness under the change of scales, and that it outperforms a state-of-the-art falsification tool.
%\keywords{signal temporal logic, cyber-physical systems, falsification, Boolean combination, reinforcement learning}
\end{abstract}

\section{Introduction}\label{sec:intro}

\paragraph{Hybrid System Falsification}
Quality assurance of \emph{cyber-physical systems (CPS)} is attracting growing attention from both academia and industry, not only because it is challenging and scientifically interesting, but also due to the safety-critical nature of many CPS. The combination of physical systems (with continuous dynamics) and digital controllers (that are inherently discrete) is referred to as \emph{hybrid systems}, capturing an important aspect of CPS. To verify hybrid systems is intrinsically hard, because the continuous dynamics therein leads to infinite search spaces.

More researchers and practitioners are therefore turning to \emph{optimization-based falsification} as a quality assurance measure for CPS.
The problem is formalized as follows. 
\begin{center}
 \begin{minipage}{0.7\textwidth}
 \underline{\bfseries The falsification problem}
 \begin{itemize}
 \item{\textbf{Given:}} 
  a \emph{model} $\mathcal{M}$ (that takes an input signal $\bu$
  and yields an output signal $\mathcal{M}(\bu)$), and
  a \emph{specification} $\varphi$ (a temporal formula)
 \item{\textbf{Find:}} 
  a \emph{falsifying input}, that is, an input signal $\bu$ such
  that the corresponding output $\mathcal{M}(\bu)$ violates $\varphi$ 
 \end{itemize}
 \end{minipage}
\begin{math}
  	\xymatrix@1@+0.8em{
 	 {}
  	 \ar[r]^-{\bu}
	  &
	  { \hspace{-0.8em} \quad\xybox{ *+++++++[F]{\mathcal{M}} }}
	   \ar[r]^-{\mathcal{M}(\bu)}_-{\not\models\varphi \; ?}
		  &
 	 {}
	  }
 \end{math}
\end{center}
In optimization-based falsification, the above problem is turned into an optimization problem. It is \emph{robust semantics} of temporal formulas~\cite{FainekosP09,DonzeM10} that makes it possible. Instead of the Boolean satisfaction relation $\bv\models\varphi$, robust semantics assigns a quantity $\sem{\bv,\varphi}\in\R\cup\{\infty,-\infty\}$ that tells us, not only whether $\varphi$ is true or not (by the sign), but also \emph{how robustly} the formula is true or false. This allows one to employ hill-climbing optimization: we iteratively generate input signals, in the direction of decreasing robustness, hoping that eventually we hit negative robustness.

\begin{table}[!tb]\centering
\auxproof{ \scalebox{.8}{\begin{tikzpicture}[auto, semithick,remember picture,
   block/.style={rectangle, draw,
        minimum width=4em, text centered, rounded corners, minimum
        height=3em,text width=6em},
    nrblock/.style={rectangle, draw,
        minimum width=5em, text centered, rounded corners, minimum
        height=3em,text width=5em}
    ]
    %\fill[color=lightgray] (2,1) -- (5.5,1) -- (5.5,-5.3) -- (2,-5.3) -- cycle;
    %\draw[color=lightgray,very thick,dashed] (2.7,1) -- (2.7,-5);
    \node[block,text width=5em](n1){the system model $\mathcal{M}$};
    \node[block,right=8em of n1](n2){STL robust semantics};
    \node[block,right=6em of n2](n3){Hill-climbing optimization};
    \coordinate[left= 1em of n1](empty0);
    \coordinate[right= 1em of n3](empty4);
    \coordinate[below=3em of empty0](empty10);
    \coordinate[below=3em of empty4](empty14);
     \draw[->] (n3) -- (empty4) -- (empty14) --
           node[auto=left] {input signal $\bu$} 
      (empty10) -- (empty0) -- (n1);
    \draw[->] (n1) -- node[auto=left, text width=6em] {output signal} node[auto=right] {$\mathcal{M}(\bu)$} (n2);
    \draw[->] (n2) -- node[auto=left] {robustness} node[auto=right] {$\sem{\mathcal{M}(\bu),\varphi}$}(n3);
\end{tikzpicture}
}
\caption{Falsification by hill-calming optimization: a workflow. Exists once $\sem{\mathcal{M}(\bu),\varphi}<0$, i.e.\ $\mathcal{M}(\bu)\not\models\varphi$}
\label{fig:workflow}
}

\caption{Boolean satisfaction $\bw\models\varphi$, and quantitative robustness values $\sem{\bw,\varphi}$, of three signals of $\speed$ for the STL formula $\varphi\equiv\BoxOp{[0,30]}(\speed < 120)$}
\label{table:exampleRobustness}
\footnotesize
\begin{tabular}{c||c|c|c}
%\hline
 signal $\bw$  &\begin{minipage}{0.25\textwidth}\vspace{+0.2em}\includegraphics[width=0.99\textwidth]{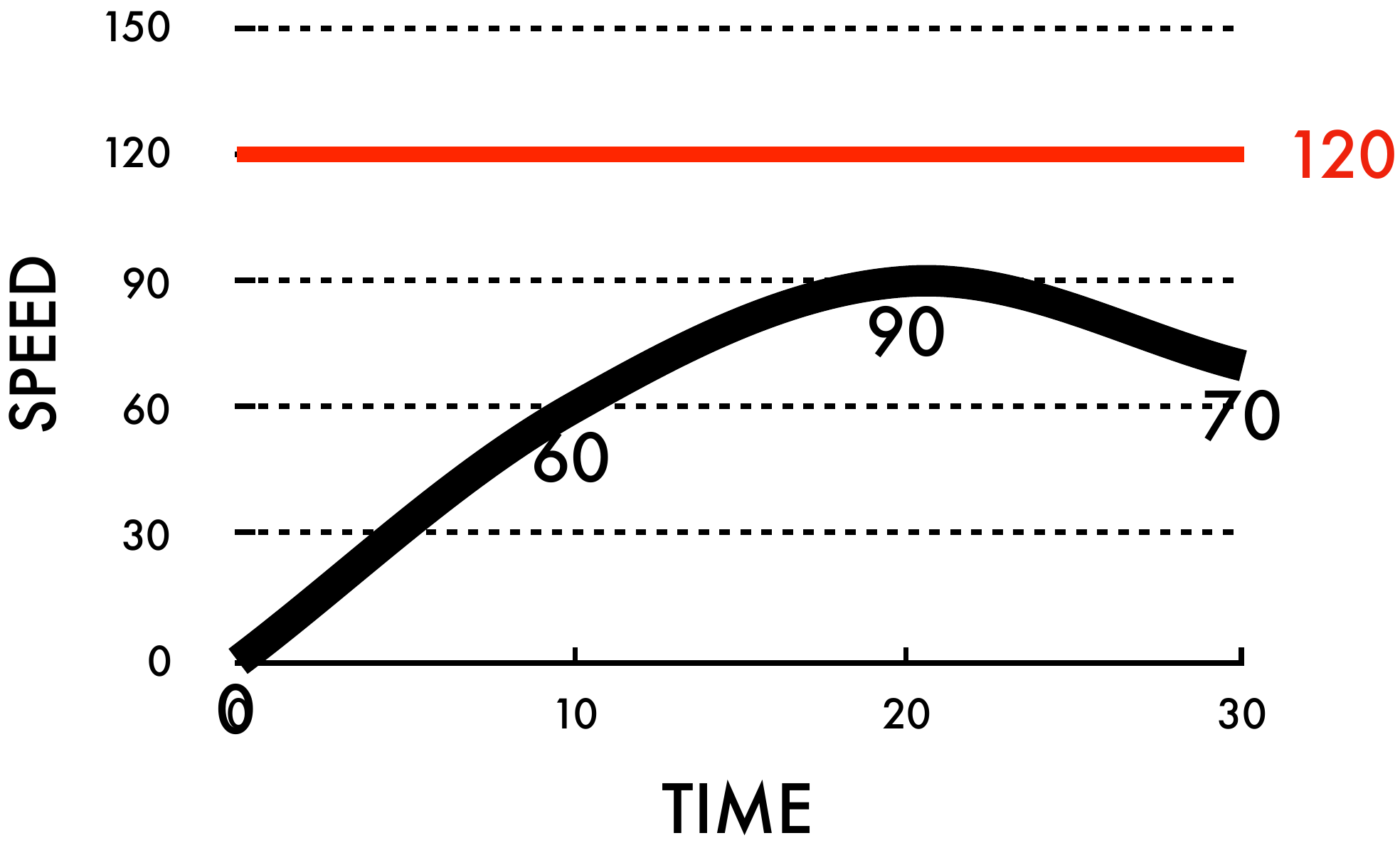} \end{minipage}\vspace{+0.2em}
 &\begin{minipage}{0.25\textwidth}\vspace{+0.2em}\includegraphics[width=0.99\textwidth]{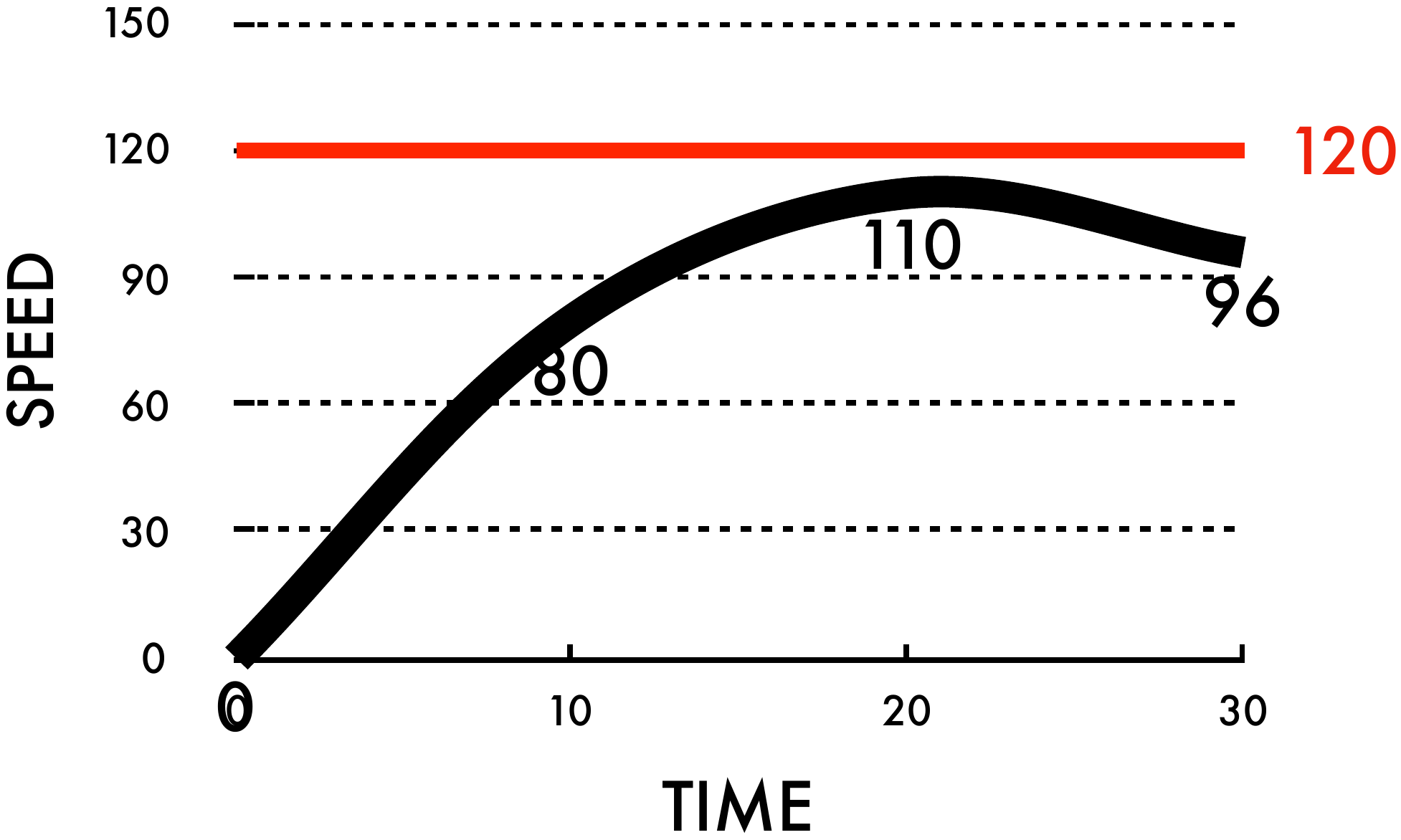} \end{minipage}  \vspace{+0.2em}
 &\begin{minipage}{0.25\textwidth}\vspace{+0.2em}\includegraphics[width=0.99\textwidth]{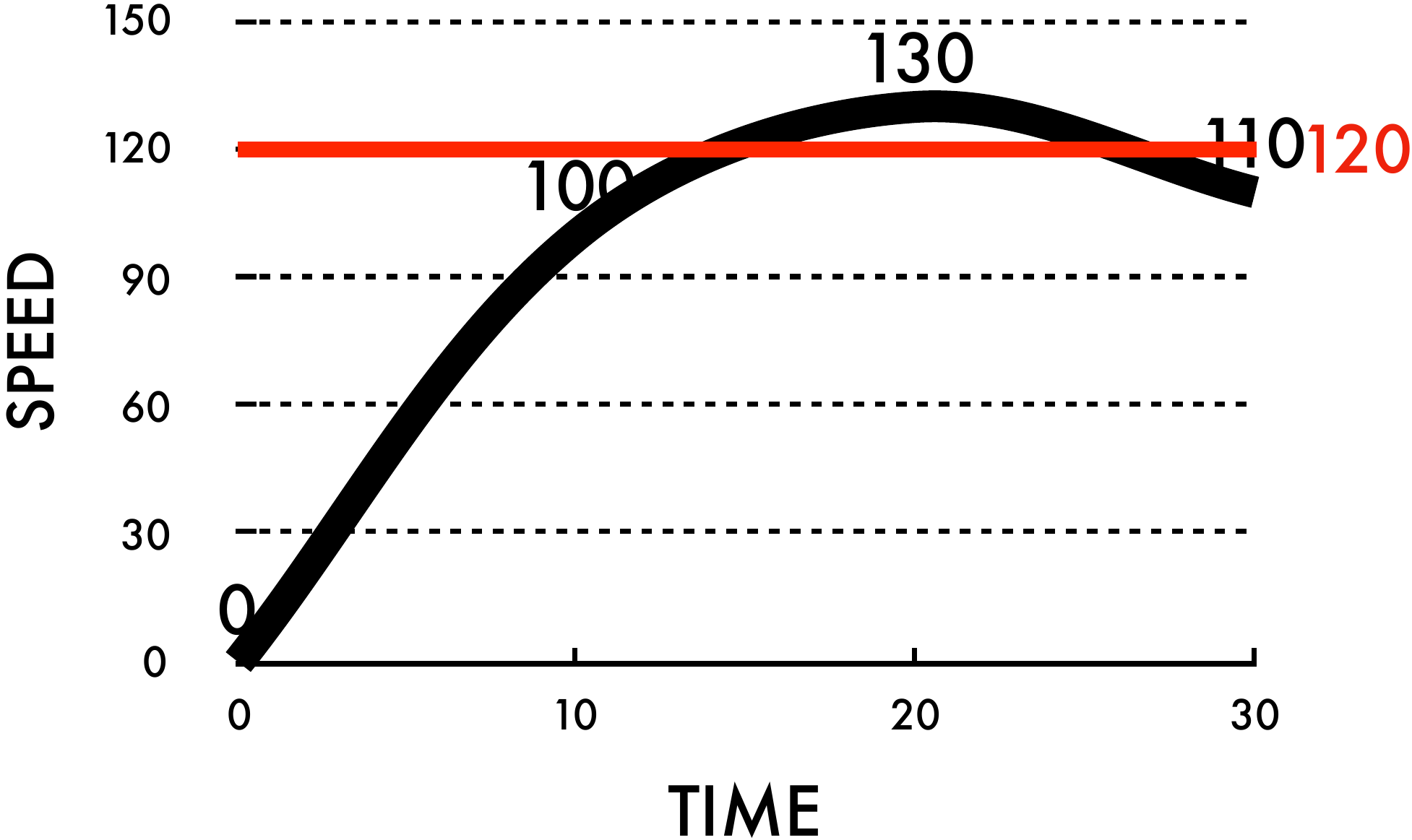} \end{minipage}\vspace{+0.2em} \\ \hline
 %Boolean\\satisfaction
$\bw\models\varphi$ 
& True & True & False \\ \hline
%quantitative\\robustness
$\sem{\bw,\varphi}$
& 30 & 10 & $-10$
%\\ \hline
\end{tabular}
\end{table}

An illustration of robust semantics is in Table~\ref{table:exampleRobustness}. We use 
\emph{signal temporal logic (STL)}~\cite{DonzeM10}, a temporal logic that is commonly used in hybrid system specification. The specification says the speed must always be below $120$ during the time interval $[0,30]$. In the search of an input signal $\bu$ (e.g.\ of throttle and brake) whose corresponding output $\M(\bu)$ violates the specification, the quantitative robustness $\sem{\M(\bu),\varphi}$ gives much more information than the Boolean satisfaction $\M(\bu)\models\varphi$. Indeed, in Table~\ref{table:exampleRobustness}, while Boolean satisfaction fails to discriminate the first two signals, the quantitative robustness indicates a tendency that the second signal is closer to violation of the specification. 

In the falsification literature, stochastic algorithms are used for hill-climbing optimization. Examples include simulated annealing (SA), globalized Nelder-Mead (GNM~\cite{LuersonLeRiche2004}) and covariance matrix adaptation evolution strategy (CMA-ES~\cite{AugerH05}). Note that the system model $\mathcal{M}$ can be black-box: we have only to observe the correspondence between input $\bu$ and output $\mathcal{M}(\bu)$. Observing an error $\mathcal{M}(\bu')$ for some input $\bu'$ is sufficient evidence for a system designer to know that the system needs improvement. Besides these practical advantages, optimization-based falsification is an interesting scientific topic: it combines two different worlds of formal reasoning and stochastic optimization.

Optimization-based falsification started in~\cite{FainekosP09} and has been developed vigorously~\cite{Annpureddy-et-al2011,AdimoolamDDKJ17,DeshmukhJKM15,KuratkoR14,Donze10,DonzeM10,DreossiDDKJD15,ZutshiDSK14,AkazakiKH17,SilvettiPB17,DreossiDS17,falsificationTCAD2018,AkazakiLYDH18,KatoIH18}. See~\cite{KapinskiDJIB16} for a survey. There are mature tools such as Breach~\cite{Donze10} and S-Taliro~\cite{Annpureddy-et-al2011}; they work with industry-standard Simulink models.

\paragraph{Challenge: The Scale Problem in Boolean Superposition}
In the field of hybrid falsification---and more generally in search-based testing---the following problem is widely recognized. We shall call the problem \emph{the scale problem (in Boolean superposition)}. 

Consider an STL specification
\begin{math}
 \varphi\;\equiv\;\BoxOp{[0,30]}(\neg( \rpm > 4000) \vee (\speed > 20))
\end{math}
for a car; it is equivalent to 
$\BoxOp{[0,30]}((\rpm > 4000) \rightarrow (\speed > 20))$ and says that 
the speed should not be too small whenever the rpm is over $4000$. According to the usual definition in the literature~\cite{FainekosP09,Donze10}, the Boolean connectives $\lnot$ and $\lor$ are interpreted by $-$ and the supremum $\sqcup$, respectively; and the ``always'' operator $\BoxOp{[0,30]}$ is by infimum $\bigsqcup$. Therefore the robust semantics of $\varphi$ under the signal $(\rpm,\speed)$, where $\rpm,\speed\colon [0,30]\to \R$, is given as follows. 
\begin{equation}\label{eq:robustnessExampleIntro}
 \sem{(\rpm,\speed), \varphi}
 = \textstyle
 \bigsqcap_{t\in [0,30]}
 \left(\,
 \bigl(4000 - \rpm(t)\bigr)
 \sqcup
 \bigl(\speed(t)-20\bigr)
\,\right)
\end{equation}
A problem is that, in the supremum of two real values in~(\ref{eq:robustnessExampleIntro}), one component can totally \emph{mask} the contribution of the other. In this specific example, the former ($\rpm$) component can have values as big as thousands, while the latter ($\speed$) component will be in the order of tens. This means that in hill-climbing optimization it is hard to use the information of both signals, as one will be masked.

Another related problem is that the efficiency of a falsification algorithm would depend on the choice of units of measure. Imagine replacing rpm with rph in~(\ref{eq:robustnessExampleIntro}), which makes the constant 4000 into 240000, and make the situation even worse. 

These problems---that we call the \emph{scale problem}---occur in many falsification examples, specifically when a specification involves Boolean connectives. We do need Boolean connectives in specifications: for example, many real-world specifications in industry are of the form 
$\BoxOp{I}(\varphi_{1}\rightarrow\varphi_{2})$, 
requiring that
an event $\varphi_{1}$ triggers a countermeasure $\varphi_{2}$ all the time.

One could use different operators for interpreting Boolean connectives. For example, in~\cite{DBLP:conf/cav/FuS16}, $\lor$ and $\land$ are interpreted by $+$ and $\times$ over $\R$, respectively. However, these choices do not resolve the scale problem, either. In general, it does not seem easy to come up with a fixed set of operators over $\R$ that interpret Boolean connectives and are free from the scale problem.

\begin{wrapfigure}[7]{r}{0.35\textwidth}
\centering

\vspace{-3.5em}
\begin{tabular}{ccc}
 \includegraphics[width=4em]{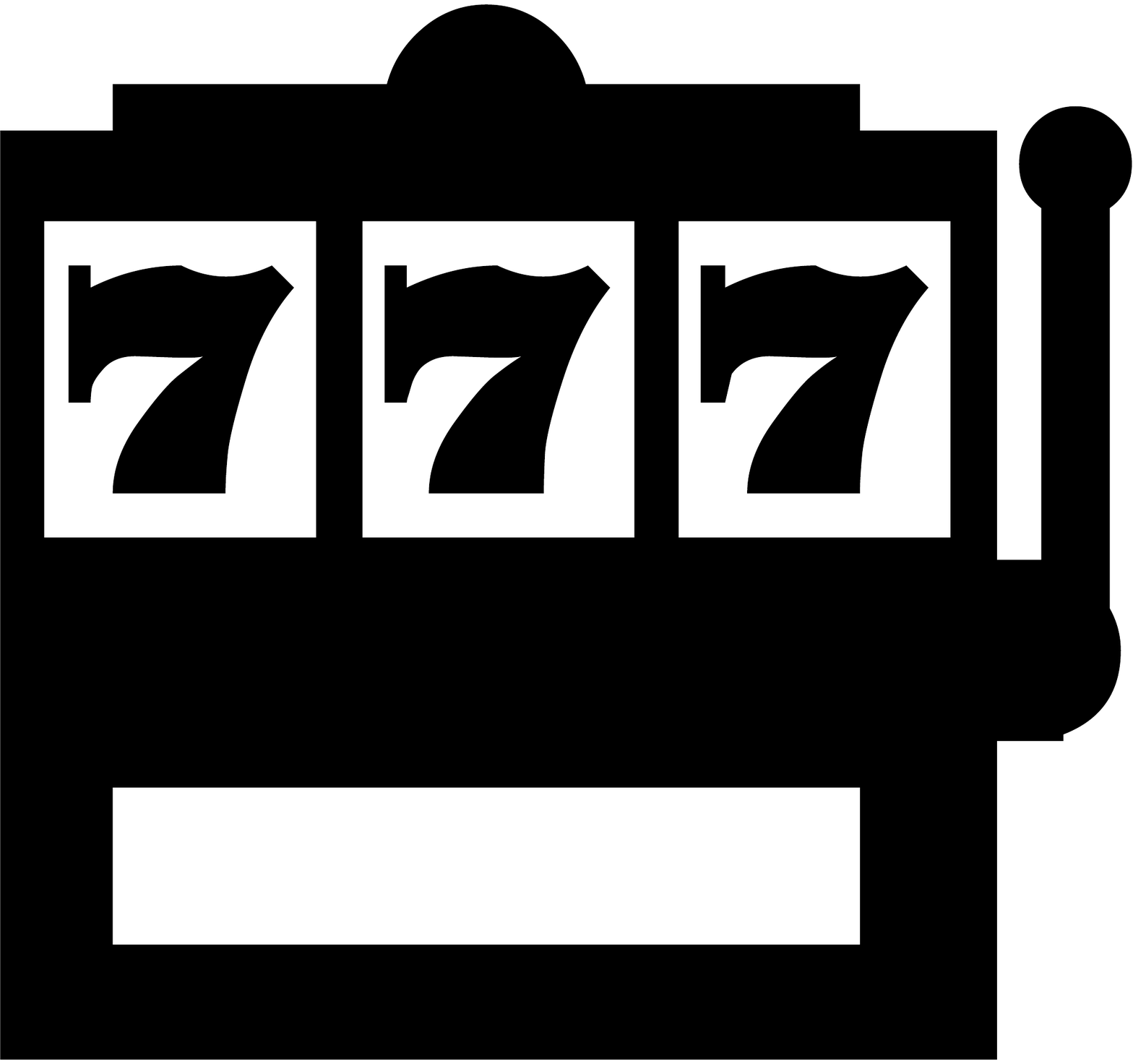}
 &\quad&
 \includegraphics[width=4em]{slot-machine-silhouette.pdf}
 \\[-1.5em]
 $\varphi_{1}$
 &&
 $\varphi_{2}$
\end{tabular}

\caption{A multi-armed bandit for falsifying $\BoxOp{I}{(\varphi_{1}\land \varphi_{1})}$
}
\label{fig:MABIntro}
\end{wrapfigure}
\paragraph{Contribution: Integrating
Multi-Armed Bandits into Optimization-Based Falsification
}
As a solution to the scale problem in Boolean superposition that we just described, we introduce a new approach that does \emph{not} superpose robustness values. Instead, we integrate \emph{multi-armed bandits (MAB)} in the existing framework of falsification guided by hill-climbing optimization. 

The MAB problem is a prototypical reinforcement learning problem: a gambler sits in front of a row of slot machines; their performance (i.e.\ average reward) is not known; the gambler plays a machine in each round and he continues with many rounds; and the goal is to optimize cumulative rewards. The gambler needs to play different machines and figure out their performance, at the cost of the loss of opportunities in the form of playing suboptimal machines. 

In this paper, we focus on specifications of the form $\BoxOp{I}(\varphi_{1}\land\varphi_{2})$ and $\BoxOp{I}(\varphi_{1}\lor\varphi_{2})$; we call them \emph{(conjunctive/disjunctive) safety properties}. We identify an instance of the MAB problem in the choice of the formula (out of $\varphi_{1},\varphi_{2}$) to try to falsify by hill climbing. See Fig.~\ref{fig:MABIntro}. We combine MAB algorithms (such as $\varepsilon$-greedy and UCB1, see~\S{}\ref{subsec:MAB}) with hill-climbing optimization, for the purpose of coping with the scale problem in Boolean superposition. This combination is made possible by introducing a novel reward notion for MAB, called \emph{hill-climbing gain}, that is tailored for this purpose.

We have implemented our MAB-based falsification framework in MATLAB, building on Breach~\cite{Donze10}.\footnote{Code obtained at \url{https://github.com/decyphir/breach}.} Our experiments with benchmarks from~\cite{HoxhaAF14,JinDKUB14,beale1992neural} demonstrate that our MAB-based approach is a viable one against the scale problem. In particular, our approach is observed to be (almost totally) robust under the change of scaling (i.e.\ changing units of measure, such as from rpm to rph that we discussed after the formula~(\ref{eq:robustnessExampleIntro})). Moreover, for the benchmarks taken from the previous works---they do not suffer much from the scale problem---our algorithm performs better than the state-of-the-art falsification tool \breach~\cite{Donze10}.

\paragraph{Related Work} 
Besides those we mentioned, we shall discuss some related works.

Formal verification approaches to correctness of hybrid systems employ a wide range of techniques, including model checking, theorem proving, rigorous numerics, nonstandard analysis, and so on~\cite{ChenAS13,GaoAC12,FrehseGDCRLRGDM11,FanQM0D16,DreossiDP16,DBLP:books/sp/Platzer18,HasuoS12CAV,DBLP:conf/icfem/LiebrenzHG18}. These are currently not very successful in dealing with complex real-world systems, due to issues like scalability and black-box components.

Our use of MAB in falsification exemplifies the role of the \emph{exploration-exploitation trade-off}, the core problem in reinforcement learning. The trade-off has been already discussed in some works on falsification. A recent example is~\cite{falsificationTCAD2018}, where they use Monte Carlo tree search to force systematic exploration of the space of input signals. Besides MCTS, \emph{Gaussian process learning (GP learning)} has also attracted attention in machine learning as a clean way of balancing exploitation and exploration. The GP-UCB algorithm is a widely used strategy there. Its use in hybrid system falsification is pursued e.g.\ in~\cite{AkazakiKH17,SilvettiPB17}.

More generally, \emph{coverage-guided falsification}~\cite{KuratkoR14,DreossiDDKJD15,DeshmukhJKM15,AdimoolamDDKJ17} aims at coping with the exploration-exploitation trade-off. One can set the current work in this context---the difference is that we force systematic exploration on the specification side, not in the input space.

There have been efforts to enhance expressiveness of MTL and STL, so that engineers can express richer intentions---such as time robustness and frequency---in specifications~\cite{AkazakiH15,NguyenKJDBJ17}. This research direction is orthogonal to ours; we plan to investigate the use of such logics in our current framework.

A similar masking problem around Boolean connectives is discussed in~\cite{DokhanchiYHF17,FerrereNDIK19}. Compared to those approaches, our technique does not need the explicit declaration of {\it input vacuity} and {\it output robustness}, but it relies on the ``hill-climbing gain'' reward to learn the significance of each signal.

Finally, the interest in the use of deep neural networks is rising in the field of falsification (as well as in many other fields). See e.g.~\cite{AkazakiLYDH18,KatoIH18}.

%Schematic overview: optimization based falsification
\section{Preliminaries: Hill Climbing-Guided Falsification}\label{sec:overview}
We review a well-adopted methodology for hybrid system falsification, namely the one guided by hill-climbing optimization. It makes essential use of quantitative \emph{robust semantics} of temporal formulas, which we review too. 

\subsection{Robust Semantics for STL}
Our definitions here are taken from~\cite{FainekosP09,DonzeM10}. 
\begin{mydefinition}[(time-bounded) signal]\label{def:timeBoundedSignal}
 Let $T\in \Rpos$ be a positive real. 
 An \emph{$M$-dimensional signal} with a time horizon $T$ is a function $\bw\colon [0,T]\to\R^{M}$.

Let $\bw\colon [0,T]\to \RM$ and $\bw'\colon [0,T']\to\RM$ be $M$-dimensional signals. Their \emph{concatenation} $\bw\cdot\bw'\colon [0,T+T']\to \RM$ is the $M$-dimensional signal defined by
 \begin{math}
 (\bw\cdot\bw')(t)=
 \bw(t)
 \end{math}
 if $t\in [0,T]$, and
  $ (\bw\cdot\bw')(t)=\bw'(t-T)$ if $t\in(T,T+T']$.

Let $0<T_{1}<T_{2}\le T$. The \emph{restriction} 
 $\bw|_{[T_{1},T_{2}]}\colon [0,T_{2}-T_{1}]\to \RM$ of $\bw\colon [0,T]\to \RM$ to the interval $[T_{1},T_{2}]$ is defined by $(\bw|_{[T_{1},T_{2}]})(t)=\bw(T_{1}+t)$. 
\end{mydefinition}
One main advantage of optimization-based falsification is that a system model can be a black box---observing the correspondence between input and output suffices. We therefore define a system model simply as a function.
% This enables the application of the methodology to models written in Simulink (which is industy-standard). 
\begin{mydefinition}[system model $\M$]\label{def:systemModel}
 A \emph{system model}, with $M$-dimensional input and $N$-dim.\ output, is a function $\mathcal{M}$ that takes 
an input signal $\bu\colon [0,T]\to \R^{M}$ and returns a signal $\mathcal{M}(\bu)\colon [0,T]\to \R^{N}$. Here the common time horizon $T\in \Rpos$ 
%of $\bu$ and $\mathcal{M}(\bu)$ 
is arbitrary. 
Furthermore, we impose the following \emph{causality} condition on $\mathcal{M}$:
for any time-bounded signals $\bu\colon [0,T]\to \R^{M}$ and $\bu'\colon [0,T']\to \R^{M}$, we require that
\begin{math}\label{eq:causality}
 \mathcal{M}(\bu\cdot\bu')
\big|_{[0,T]}
= \mathcal{M}(\bu)
\end{math}. 

\end{mydefinition}

\begin{mydefinition}[STL syntax]\label{def:stlSyntax}
We fix a set $\Var$ of variables. In $\STL$, 
 %the set $\AP$ of
 \emph{atomic propositions} and 
%the set
% $\Fml$
% of
 \emph{formulas} are defined as follows, respectively:
\begin{math}
%    \AP \ni 
      \alpha 
 \,::\equiv\,
        f(x_1, \dots, x_N) > 0
\end{math}, and 
\begin{math}
%      \Fml \ni 
      \varphi  \,::\equiv\,
        \alpha \mid \bot
        % \mid \top
        \mid \neg \varphi 
        \mid \varphi \wedge \varphi 
        \mid \varphi \vee \varphi 
        % \mid \varphi \wedge \varphi
        \mid \varphi \UntilOp{I} \varphi
        % \mid \varphi \TUntil{[a,b]} \varphi 
        % \mid \varphi \LTUntil{[a,b]} \varphi 
\end{math}. Here
 $f$ is an $N$-ary function $f:\RN \to \R$, $x_1, \dots, x_N \in \Var$,
%  $a,b \in \Rnn$ such that $a<b$,
  and $I$ is a closed non-singular interval in $\Rnn$,
  i.e.\ $I=[a,b]$ or $[a, \infty)$ where $a,b \in \R$ and $a<b$.
% Eventually operator $\DiaOp{I}\varphi$ can be derived by $\DiaOp{I}{\varphi} ::\equiv \top\UntilOp{I} \varphi$; and always operator $\BoxOp{I}{\varphi}$
% can be derived by $\BoxOp{I}{\varphi} ::\equiv \neg \DiaOp{I}{\neg\varphi}$.
\end{mydefinition}
 We omit subscripts $I$ for temporal operators if $I = [0, \infty)$. Other common connectives such as $\rightarrow,\top$, $\Box_{I}$ (always) and $\Diamond_{I}$ (eventually), are introduced as abbreviations: $\Diamond_{I}\varphi\equiv\top\UntilOp{I}\varphi$ and 
$\Box_{I}\varphi\equiv\lnot\Diamond_{I}\lnot\varphi$. An atomic formula $f(\vec{x})\le c$, where $c\in\R$, is  accommodated using $\lnot$ and the function $f'(\vec{x}):=f(\vec{x})-c$.

%The definition of STL robustness semantics 
\begin{mydefinition}[robust semantics~\cite{DonzeM10}]\label{def:robSemantics} 
Let $\bw \colon [0,T]\to \R^{N}$ be an $N$-dimensional signal, 
and $t\in [0,T)$.
The \emph{$t$-shift} of $\bw$, denoted by $\bw^t$, is the time-bounded signal $\bw^t\colon [0,T-t]\to \R^{N}$ defined by
 $\bw^t(t') \Defeq \bw(t+t')$.

Let $\bw \colon [0,T] \to \R^{|\Var|}$ be a signal,
  and $\varphi$ be an $\STL$ formula.
  We define the \emph{robustness} 
  $\Robust{\bw}{\varphi} \in \R \cup \{\infty,-\infty\}$ as follows, by induction on the construction of formulas.
  Here $\bigsqcap$ and $\bigsqcup$ denote infimums and supremums of real numbers, respectively. Their binary version $\sqcap$ and $\sqcup$ denote minimum and maximum.
\begin{align}
\nonumber
&\Robust{\bw}{f(x_1, \cdots, x_n) > 0}  \;\Defeq \;
f\bigl(\bw(0)(x_1), \cdots, \bw(0)(x_n)\bigr) 
\\
\nonumber
&\Robust{\bw}{\bot}  \;\Defeq\;  -\infty
\qquad
\Robust{\bw}{\neg \varphi}   \;\Defeq\;   - \Robust{\bw}{\varphi}\qquad 
\\\label{eq:robustnessForBinaryBooleanOp}
&
          \Robust{\bw}{\varphi_1 \wedge \varphi_2}   \;\Defeq\;   \Robust{\bw}{\varphi_1} \sqcap \Robust{\bw}{\varphi_2}
\qquad
          \Robust{\bw}{\varphi_1 \lor \varphi_2}   \;\Defeq\;   \Robust{\bw}{\varphi_1} \sqcup \Robust{\bw}{\varphi_2}
\\
\nonumber
&
          \Robust{\bw}{\varphi_1 \UntilOp{I} \varphi_2}   \;\Defeq\; 
                                                         \textstyle{ \Vee{t \in I\cap [0,T]}\bigl(\,\Robust{\bw^t}{\varphi_2} \sqcap 
                                                         \Wedge{t' \in [0, t)} \Robust{\bw^{t'}}{\varphi_1}\,\bigr)}
    \end{align}
\end{mydefinition}

For atomic formulas,
 $\Robust{\bw}{f(\vec{x})>c}$ stands for the vertical margin $f(\vec{x})-c$ for the signal $\bw$ at time $0$.  A negative robustness value indicates how far the formula is from being true. 
It follows 
from the definition
 that the robustness for the eventually modality is given by
\begin{math}
\Robust{\bw}{\DiaOp{[a,b]} (x > 0)}
      = \Vee{t \in [a,b]\cap[0,T]} \bw(t)(x)
\end{math}. 

The above robustness notion taken from~\cite{DonzeM10} is therefore \emph{spatial}. Other robustness notions take \emph{temporal} aspects into account, too, such as ``how long before the deadline the required event occurs.'' See e.g.~\cite{DonzeM10,AkazakiH15}. Our choice of spatial robustness in this paper is for the sake of simplicity, and is thus not essential.

The original semantics of $\STL$ is Boolean, given as usual by a binary relation $\models$ between signals and formulas. The robust semantics refines the Boolean one in the following sense:
$ \sem{\bw,\varphi} > 0$
 implies
$\bw\models\varphi$, and
$ \sem{\bw,\varphi} < 0$
implies
$\bw\not\models\varphi$, see~\cite[Prop.~16]{FainekosP09}.
Optimization-based falsification via robust semantics hinges on this refinement.

\subsection{Hill Climbing-Guided Falsification}\label{subsec:hillClimbingFalsification}
As we discussed in the introduction, the falsification problem attracts growing industrial and academic attention. Its solution methodology by hill-climbing optimization is an established field, too: see~\cite{Annpureddy-et-al2011,AdimoolamDDKJ17,DeshmukhJKM15,KuratkoR14,Donze10,DonzeM10,DreossiDDKJD15,ZutshiDSK14,AkazakiKH17,SilvettiPB17,DreossiDS17,KapinskiDJIB16} and the tools Breach~\cite{Donze10} and S-TaLiRo~\cite{Annpureddy-et-al2011}. We formulate the problem and the methodology, for later use in describing our multi-armed bandit-based algorithm.

\begin{mydefinition}[falsifying input]\label{def:falsifyingInput}
Let $\mathcal{M}$ be a system model, and~$\varphi$ be an STL formula. A signal $\bu\colon [0,T]\to \R^{|\Var|}$ is a \emph{falsifying input} if $\Robust{\mathcal{M}(\bu)}{\varphi}<0$; the latter implies $\mathcal{M}(\bu)\not\models\varphi$.
\end{mydefinition}
The use of quantitative robust semantics $\Robust{\mathcal{M}(\bu)}{\varphi}\in \R\cup \{\infty,-\infty\}$ in the above problem enables the use of hill-climbing optimization.

\begin{mydefinition}[hill climbing-guided falsification]\label{def:hillClimbGuidedFalsification}
Assume the setting in Def.~\ref{def:falsifyingInput}. For finding a falsifying input, the methodology of \emph{hill climbing-guided falsification} is presented in Algorithm~\ref{algo:hillClimbingFalsification}.

Here the function $\textsc{Hill-Climb}$ makes a guess of an input signal $\bu_{k}$, aiming at minimizing the robustness $\Robust{\mathcal{M}(\bu_{k})}{\varphi}$. It does so, learning from the previous  observations 
\begin{math}
 \bigl(\,
  \bu_{l}, \,
  \Robust{\mathcal{M}(\bu_{l})}{\varphi}
\,\bigr)_{l\in [1,k-1]}
\end{math} of  input signals $\bu_{1},\dotsc, \bu_{k-1}$ and their corresponding robustness values (cf.\ Table~\ref{table:exampleRobustness}). 
\end{mydefinition}
The $\textsc{Hill-Climb}$ function can be implemented by various stochastic optimization algorithms. Examples are  CMA-ES~\cite{AugerH05} (used in our experiments),  SA, and GNM~\cite{LuersonLeRiche2004}.

\begin{algorithm}[tbp]
\caption{Hill climbing-guided falsification}
\label{algo:hillClimbingFalsification}
\begin{algorithmic}[1]
\Require a system model $\M$, an STL formula $\varphi$, 
%an initial input signal $\bu_{0}$ (that is usually chosen arbitrarily), 
and a budget $K$
\Function{Hill-Climb-Falsify}{$\M,\varphi,K$}
%\State $\vec \bu \gets \vec \bu_0$
\State $\mathsf{rb} \gets \infty$\;; \quad $k\gets 0$
   \Comment{$\mathsf{rb}$ is the smallest robustness so far, initialized to $\infty$}
\While{$\mathsf{rb}\ge 0$ and $k\le K$}
        \State $k\gets k+1$
	\State $\bu_{k}\gets$\,
         \textsc{Hill-Climb}{$\Bigl(\,
		       \bigl(\,
		       \bu_{l}, \,
		       \Robust{\mathcal{M}(\bu_{l})}{\varphi}
		       \,\bigr)_{l\in [1,k-1]}\,\Bigr)
		      $}
	\State $\mathsf{rb}_{k}\gets \sem{\M(\bu_{k}), \varphi}$
	%\State $i\gets B$
	\IIf{$\mathsf{rb}_{k} < \mathsf{rb}$} $\mathsf{rb} \gets \mathsf{rb}_{k}$ \EndIIf
\EndWhile%{$R\ge 0$ and within the budget}

\vspace{.3em}
\State$\bu\gets
\begin{cases}
 \bu_{k} & \text{if $\mathsf{rb}<0$, that is, $\mathsf{rb}_{k}=\sem{\M(\bu_{k}), \varphi}<0$}
\\
 \text{Failure}  & \text{otherwise, that is, no falsifying input found within budget $K$}
\end{cases}
$
\State \Return{$\bu$}
\EndFunction
\end{algorithmic}
\end{algorithm}

\section{Our Multi-Armed Bandit-Based Falsification Algorithm}\label{sec:mabrob}
In this section, we present our contribution, namely a falsification algorithm that addresses the scale problem in Boolean superposition  (see~\S{}\ref{sec:intro}). The main novelties in the  algorithm are as follows.
\begin{compactenum}
 \item \textbf{(Use of MAB algorithms)}
 For binary Boolean connectives, unlike most works in the field, we do not superpose the robustness values of the constituent formulas $\varphi_{1}$ and $\varphi_{2}$ using a fixed operator (such as $\sqcap$ and $\sqcup$ in~(\ref{eq:robustnessForBinaryBooleanOp})). Instead, we view the situation as an instance of the multi-armed bandit problem (MAB): we use an algorithm for MAB to choose one formula $\varphi_{i}$ to focus on (here $i\in \{1,2\}$); and then we apply hill climbing-guided falsification to the chosen formula $\varphi_{i}$. 
 \item \textbf{(\emph{Hill-climbing gain} as rewards in MAB)} For our integration of MAB  and hill-climbing optimization, the technical challenge is find a suitable notion of reward for MAB. We introduce a novel notion that we call \emph{hill-climbing gain}: it formulates the (downward) robustness gain that we would obtain by applying hill-climbing optimization, suitably normalized using the scale of previous robustness values. 
\end{compactenum}
Later, in~\S{}\ref{sec:experiments}, we demonstrate that combining those two features gives rise to falsification algorithms that successfully cope with the scale problem in Boolean superposition. 

Our algorithms focus on a fragment of STL as target specifications. They are called \emph{(disjunctive and conjunctive) safety properties}. In~\S{}\ref{subsec:lifeLongProp} we describe this fragment of STL,  and introduce necessary adaptation of the semantics. After reviewing the MAB problem in~\S{}\ref{subsec:MAB}, we present our algorithms in~\S{}\ref{subsec:ourAlgorithm1}--\ref{subsec:ourAlgorithm2}.

\subsection{Conjunctive and Disjunctive Safety Properties}\label{subsec:lifeLongProp}
\begin{mydefinition}[conjunctive/disjunctive safety property]\label{def:conjDisjSafetyProp}
An STL formula of the form $\BoxOp{I}(\varphi_{1}\land\varphi_{2})$ is called a \emph{conjunctive safety property}; an STL formula of the form
 $\BoxOp{I}(\varphi_{1}\lor\varphi_{2})$ is called a \emph{disjunctive safety property}. 
\end{mydefinition}

It is known that, in industry practice, a majority of specifications is of the form $\BoxOp{I}(\varphi_{1}\rightarrow\varphi_{2})$, where $\varphi_{1}$ describes a trigger and $\varphi_{2}$ describes a countermeasure that should follow. This property is equivalent to $\BoxOp{I}(\lnot \varphi_{1}\lor\varphi_{2})$, and is therefore a disjunctive safety property. 

In~\S{}\ref{subsec:ourAlgorithm1}--\ref{subsec:ourAlgorithm2}, we present two falsification algorithms, for conjunctive and disjunctive safety properties respectively. For the reason we just discussed, we expect the disjunctive algorithm should be more important in real-world application scenarios. In fact, the disjunctive algorithm turns out to be more complicated, and it is best introduced as an extension of the conjunctive algorithm.

We define the restriction of robust semantics to a (sub)set of time instants. 
%We will use it in the disjunctive algorithm. 
Note that we do not require $\mathcal{S}\subseteq [0,T]$ to be a single interval.

\begin{mydefinition}[{$\sem{\bw, \psi}_{\mathcal{S}}$}, 
robustness restricted to 
{$\mathcal{S}\subseteq [0,T]$}
]
\label{def:robustnessOfSafetyPropertiesRestricted}
Let $\bw \colon [0,T]\to \R^{|\Var|}$ be a signal, $\psi$ be an STL formula, and $\mathcal{S}\subseteq [0,T]$ be a subset. 
We define the \emph{robustness} of $\bw$ under $\psi$ \emph{restricted to}   $\mathcal{S}
%\subseteq [0,T]
$ by
\begin{equation}
 \sem{\bw, \psi}_{\mathcal{S}} \;\Defeq\; 
\textstyle\bigsqcap_{t\in \mathcal{S} }{%\left(\,
%\Rho{\bw}{\psi}{t}
\sem{\bw^{t},\psi}
%\right)
}\enspace.
\end{equation}
\end{mydefinition}
Obviously, $ \sem{\bw, \psi}_{\mathcal{S}} <0$ implies that there exists $t\in \mathcal{S}$ such that $ \sem{\bw^{t}, \psi}_{\mathcal{S}} <0$. We derive the following easy lemma; it is used later in our algorithm.

\begin{mylemma}\label{lem:robustSemRestrictionToSubset}
In the setting of Def.~\ref{def:robustnessOfSafetyPropertiesRestricted}, consider a disjunctive safety property $\varphi\equiv\BoxOp{I}(\varphi_{1}\lor\varphi_{2})$, and let $\mathcal{S}\Defeq\{t\in I\cap [0,T]\mid \sem{\bw^{t}, \varphi_{1}}<0\}$. Then $\sem{\bw,\varphi_{2}}_{\mathcal{S}}<0$ implies  $\sem{\bw,\BoxOp{I}(\varphi_{1}\lor\varphi_{2})}<0$.
\qed
\end{mylemma}

\subsection{The Multi-Armed Bandit   (MAB) Problem}\label{subsec:MAB}
The \emph{multi-armed bandit}  (MAB) problem describes a situation where, 
\begin{compactitem}
 \item a gambler sits in front of 
a row $A_{1},\dotsc,A_{n}$ of slot machines;
 \item each slot machine $A_{i}$ gives, when its arm is played (i.e.\ in each attempt),  a reward according to a prescribed (but unknown) probability distribution $\mu_{i}$;
 \item and the goal is to maximize the cumulative reward after a number of attempts, playing a suitable arm in each attempt.
\end{compactitem}
The best strategy of course is to keep playing the best arm $A_{\max}$, i.e.\ the one whose average reward $\mathsf{avg}(\mu_{\max})$ is the greatest. This best strategy is infeasible, however, since the distributions $\mu_{1},\dotsc,\mu_{n}$ are initially unknown. Therefore the gambler must learn about $\mu_{1},\dotsc,\mu_{n}$ through attempts.

The MAB problem exemplifies the ``learning by trying'' paradigm of \emph{reinforcement learning}, and is thus heavily studied. The greatest challenge is to balance between \emph{exploration} and \emph{exploitation}. A greedy (i.e.\ exploitation-only) strategy will play the arm whose empirical average reward is the maximum. However, since the rewards are random, this way the gambler can miss another arm whose real performance is even better but which is yet to be found so. Therefore one needs to mix exploration, too, occasionally trying empirically non-optimal arms, in order to identity their true performance.

The relevance of MAB to our current problem is as follows. Falsifying a conjunctive safety property $\BoxOp{I}(\varphi_{1}\land\varphi_{2})$ amounts to finding a time instant $t\in I$ at which either $\varphi_{1}$ or $\varphi_{2}$ is falsified. We can see the two subformulas ($\varphi_{1}$ and $\varphi_{2}$) as two arms, and this constitutes an instance of the MAB problem. In particular, playing an arm translates to a falsification attempt by hill climbing, and collecting rewards translates to spending time to minimize the robustness. We show in \S{}\ref{subsec:ourAlgorithm1}--\ref{subsec:ourAlgorithm2} that this basic idea extends to disjunctive safety properties $\BoxOp{I}(\varphi_{1}\lor\varphi_{2})$, too.

A rigorous formulation of the MAB problem is presented  for the record. 
\begin{mydefinition}[the multi-armed bandit problem]\label{def:mabproblem}
The \emph{multi-armed bandit} (MAB) problem is formulated as follows. 

\noindent\textbf{Input:}  arms $(A_1, \dots, A_n)$,  the associated probability distributions $\mu_{1},\dotsc,\mu_{n}$ over $\R$, and a time horizon $H\in \N\cup\{\infty\}$.

\noindent\textbf{Goal:} synthesize a sequence $A_{i_{1}}A_{i_{2}}\dotsc A_{i_{H}}$, so that  the cumulative reward
\begin{math}
 \sum_{k=1}^{H}\rew_{k}
\end{math} is maximized. Here the reward $\rew_{k}$ of the $k$-th attempt is sampled from the distribution $\mu_{i_{k}}$ associated with the arm $A_{i_{k}}$ played at the $k$-th attempt.

We introduce some notations for later use. Let 
$(A_{i_{1}}\dotsc A_{i_{k}}, \rew_{1}\dotsc \rew_{k})$ be a \emph{history}, i.e.\ the sequence of arms  played so far (here $i_{1},\dotsc, i_{k}\in [1,n]$), and the sequence of rewards obtained by those attempts ($\rew_{l}$ is sampled from $\mu_{i_{l}}$). 

For an arm $A_{j}$, its  \emph{visit count} $N(j,A_{i_{1}}A_{i_{2}}\dotsc A_{i_{k}}, \rew_{1}\rew_{2}\dotsc \rew_{k})$ is given by the number of occurrences of $A_{j}$ in $A_{i_{1}}A_{i_{2}}\dotsc A_{i_{k}}$. 
%For an arm $A_{j}$, 
Its  \emph{empirical average reward} $R(j,A_{i_{1}}A_{i_{2}}\dotsc A_{i_{k}}, \rew_{1}\rew_{2}\dotsc \rew_{k})$ is given by 
\begin{math}
 \sum_{l\in \{l\in [1,k]\mid i_{l}=j\}} \rew_{l}
\end{math}, i.e.\ the average return of the arm $A_{j}$ in the history. 
When the history is obvious from the context, we simply write $N(j,k)$ and $R(j,k)$.
\end{mydefinition}

\subsubsection{MAB Algorithms}\label{subsec:algorithmsForMAB}
There have been a number of algorithms proposed for the MAB problem; each of them gives a \emph{strategy} (also called a \emph{policy}) that tells which arm to play, based on the previous attempts and their rewards. The focus here is how to resolve the exploration-exploitation trade-off. 
Here we review two well-known algorithms.

\paragraph{The $\varepsilon$-Greedy Algorithm} 
%$\varepsilon$-greedy 
This is a simple algorithm that spares a small fraction $\varepsilon$ of chances for empirically non-optimal arms. The spared probability $\varepsilon$ is uniformly distributed. 
See Algorithm~\ref{algo:greedy}. 

\begin{algorithm}[!tb]
\caption{The $\varepsilon$-greedy algorithm for multi-armed bandits}
\label{algo:greedy}
\begin{algorithmic}[1]
\Require the setting of Def.~\ref{def:mabproblem}, and a constant $\varepsilon >0$ (typically very small)
%\Statex
%\Function{Choose}{$x,y$}
\Statex 
At the $k$-th attempt, choose the arm $A_{i_{k}}$ as follows

\State $j_{\text{emp-opt}}\gets\textstyle \argmax_{j\in[1,n]}R(j,k-1)$ 
 \Comment{the arm that is empirically optimal}
\State Sample $i_{k}\in [1,n]$ from the distribution
\Statex\qquad
       \begin{math}
	\left[
         \begin{array}{rcl}
	   j_{\text{emp-opt}} &\longmapsto& (1-\varepsilon) + \frac{\varepsilon}{n}
          \\
	   j &\longmapsto& \frac{\varepsilon}{n} \qquad\text{for each $j\in [1,n]\setminus \{j_{\text{emp-opt}}\}$}
	 \end{array}
	\right]
       \end{math} 

\State \Return $i_{k}$
\end{algorithmic}
\end{algorithm}

\paragraph{The UCB1 Algorithm} The UCB1  (\emph{upper confidence bound}) algorithm is more complex; it comes with a theoretical upper bound for  \emph{regrets}, i.e.\ the gap between the expected cumulative reward and the optimal (but infeasible) cumulative reward (i.e.\ the result of 
keep playing the optimal  arm $A_{\max}$). It is known that the UCB1 algorithm's regret is at most $O(\sqrt{nH\log{H}})$ after $H$ attempts, improving the naive random strategy (which has the expected regret $O(H)$). 

See  Alg.~\ref{algo:ucb1}. The algorithm is deterministic, and picks the arm that maximizes the value shown in Line~\ref{line:UCBScore}.
%---it is called the UCB score. 
The first term $R(j,k-1)$ is the \emph{exploitation} factor, reflecting the arm's  empirical performance.  The second term is the \emph{exploration} factor. Note that it is bigger if the arm $A_{j}$ has been played less frequently. Note also that the exploration factor eventually decays over time: the denominator grows roughly with $O(k)$, while the numerator grows with $O(\ln k)$.

\begin{algorithm}[!tb]
\caption{The UCB1 algorithm for multi-armed bandits}
\label{algo:ucb1}
\begin{algorithmic}[1]
\Require the setting of Def.~\ref{def:mabproblem}, and a constant $c >0$
\Statex 
At the $k$-th attempt, choose the arm $A_{i_{k}}$ as follows
\State\label{line:UCBScore}
 $i_{k}\gets
\argmax_{j\in[1,n]}{\left(R(j,k-1) + c\sqrt{\textstyle\frac{2\ln (k-1)}{ N(j,k-1)}}\right)}$
\State \Return $i_{k}$
\end{algorithmic}
\end{algorithm}

\subsection{Our MAB-Guided  Algorithm I: Conjunctive Safety Properties}\label{subsec:ourAlgorithm1}

Our first algorithm targets at conjunctive safety properties. It is
based on our identification of MAB in a Boolean conjunction in
falsification---this is as we discussed just above
Def.~\ref{def:mabproblem}. The technical novelty lies in the way we
combine MAB algorithms and hill-climbing optimization; specifically, we
introduce the notion of \emph{hill-climbing gain} as a reward notion in
MAB (Def.~\ref{def:hillClimbGain}). This first algorithm paves the way to the one for disjunctive
safety properties, too (\S{}\ref{subsec:ourAlgorithm2}).

\begin{algorithm}[!tb]
\caption{Our MAB-guided algorithm I: \emph{conjunctive} safety properties}
\label{algo:main1}
\begin{algorithmic}[1]
\Require a system model $\M$, an STL formula $\varphi\equiv \BoxOp{I}(\varphi_{1}\land\varphi_{2})$, and a budget $K$
\Function{MAB-Falsify-Conj-Safety}{$\M,\varphi,K$}
%\State ($\bu_0 \gets$ \Call{Preprocess}{$\M, \varphi$, budget\_pre})
%\State $\bu \gets \bu_0$
 \State % $\mathsf{rb}_{1}, \mathsf{rb}_{2} \gets \infty$\;; \quad
        $\mathsf{rb}\gets \infty$\;; \quad
%       $N_{1}, N_{2} \gets 0$\;; \quad
       $k\gets 0$
 \Statex
   \Comment{$\mathsf{rb}$ is the smallest robustness seen so far, for either $\BoxOp{I}\varphi_{1}$ or $\BoxOp{I}\varphi_{2}$}
\While{$\mathsf{rb}\ge 0$
%, $\mathsf{rb}_{2}\ge 0$
   and $k\le K$}
\Comment{iterate if not yet falsified, and within budget}
\State $k\gets k+1$
\State\label{line:MABConj} $i_{k}\gets \textsc{MAB}\Bigl(\,(\varphi_{1},\varphi_{2}),\,\bigl(\mathcal{R}(\varphi_{1}), \mathcal{R}(\varphi_{2})\bigr),\, \varphi_{i_{1}}\dotsc \varphi_{i_{k-1}},
\, \rew_{1}\dotsc \rew_{k-1}\, \Bigr)$
\Statex \Comment{an MAB choice of $i_{k}\in \{1,2\}$ for optimizing the reward $\mathcal{R}(\varphi_{i_{k}})$}

 \State \label{line:HillClimbConj}
  $\bu_{k}\gets \textsc{Hill-Climb}
    \left(\,
     \bigl(\,
    (\bu_{l},\, 
    %\Robust{\mathcal{M}(\bu_{l})}{\BoxOp{I}\varphi_{i_{k}}}
    \mathsf{rb}_{l}
    )
   \,\bigr)_{l\in [1,k-1] \text{ such that } i_{l}=i_{k}}
   \,\right)
 $
 \Statex \Comment{
 \parbox[t]{.85\textwidth}{suggestion of the next input $\bu_{k}$ by hill
 climbing,  based on the previous observations on the formula
 $\varphi_{i_{k}}$ (those on the other formula are ignored)
 }}

	\State\label{line:defRbConj}
         $\mathsf{rb}_{k}\gets \sem{\M(\bu_{k}), \BoxOp{I}\varphi_{i_{k}}}$
	\IIf{$\mathsf{rb}_{k} < \mathsf{rb}$} \label{line:updateRbConj}
	  $\mathsf{rb} \gets \mathsf{rb}_{k}$
	\EndIIf

\EndWhile

\vspace{.3em}
\State$\bu\gets
\begin{cases}
 \bu_{k} & \text{if $\mathsf{rb}<0$
%, that is, $\mathsf{rb}_{k}=\sem{\M(\bu_{k}), \BoxOp{I}\varphi_{i_{k}}}<0$
}
\\
 \text{Failure}  & \text{otherwise, that is, no falsifying input found within budget $K$}
\end{cases}
$
\State \Return{$\bu$}

\EndFunction
\end{algorithmic}
\end{algorithm}

\begin{algorithm}[!tb]
\caption{Our MAB-guided algorithm II: \emph{disjunctive} safety properties}
\label{algo:main2}
\begin{algorithmic}[1]
\Require a system model $\M$, an STL formula $\varphi\equiv \BoxOp{I}(\varphi_{1}\lor\varphi_{2})$, 
and a budget $K$
\Function{MAB-Falsify-Disj-Safety}{$\M,\varphi,K$}
\Statex The same as Algorithm~\ref{algo:main1}, except that Line~\ref{line:defRbConj} is replaced by the following Line~\ref{line:defRbConj}'.

\vspace{.3em}
\Statex \ref{line:defRbConj}': \quad 
 $\mathsf{rb}_{k}\gets \sem{\M(\bu_{k}), \varphi_{i_{k}}}_{\mathcal{S}_{k}}$
%\Statex \qquad\qquad\qquad where
 \quad where
 $\mathcal{S}_{k}=\bigl\{\,t\in I\cap [0,T]\,\big|\, \sem{\M(\bu_{k}^{t}), \varphi_{\overline{i_{k}}}}<0\,\bigr\}$
\Statex\Comment{here $\varphi_{\overline{i_{k}}}$ denotes the other formula than $\varphi_{i_{k}}$, among $\varphi_{1},\varphi_{2}$}
\EndFunction
\end{algorithmic}
\end{algorithm}

The algorithm is in Algorithm~\ref{algo:main1}.  Some remarks are in order.

Algorithm~\ref{algo:main1}  aims to falsify a conjunctive safety
property $\varphi\equiv\BoxOp{I}(\varphi_{1}\land\varphi_{2})$. Its
overall structure is to
\emph{interleave} two sequences of falsification attempts, both of which are hill
climbing-guided. These two sequences of attempts aim to falsify
$\BoxOp{I}\varphi_{1}$
and
 $\BoxOp{I}\varphi_{2}$, respectively. Note that $
\sem{\M(\bu),\varphi}\le
 \sem{\M(\bu),\BoxOp{I}\varphi_{1}}$, therefore falsification of
 $\BoxOp{I}\varphi_{1}$ implies falsification of  $\varphi$; the same holds
 for $\BoxOp{I}\varphi_{2}$, too.

 In Line~\ref{line:MABConj} we run an MAB algorithm to decide which of
 $\BoxOp{I}\varphi_{1}$  and $\BoxOp{I}\varphi_{2}$  to target at in the
 $k$-th attempt. The function $\textsc{MAB}$ takes the following as its arguments: 1) the list of arms, given by the formulas $\varphi_{1},\varphi_{2}$; 2) their rewards 
$\mathcal{R}(\varphi_{1}),
\mathcal{R}(\varphi_{2})$; 3) the history $\varphi_{i_{1}}\dotsc \varphi_{i_{k-1}}$ of previously played arms ($i_{l}\in\{1,2\}$); and 4) the history $\rew_{1}\dotsc \rew_{k-1}$ of previously observed rewards.  This way, the type of the $\textsc{MAB}$  function
in  Line~\ref{line:MABConj} 
 matches the format in Def.~\ref{def:mabproblem}, and thus the function   can be instantiated with any MAB algorithm such as Algorithms~\ref{algo:greedy}--\ref{algo:ucb1}. 

The only missing piece is the definition of the rewards $\mathcal{R}(\varphi_{1}),
\mathcal{R}(\varphi_{2})$. We introduce the following notion, tailored for combining MAB and hill climbing.

\begin{mydefinition}[hill-climbing gain]\label{def:hillClimbGain}
 In   Algorithm~\ref{algo:main1}, in Line~\ref{line:MABConj}, the reward $\mathcal{R}(\varphi_{i})$ of the arm $\varphi_{i}$ (where $i\in\{1,2\}$) is defined by
\begin{equation*}\label{eq:hillClimbingGain}
\begin{aligned}
&  \MR(\varphi_{i}) = 
 \begin{cases}
 \displaystyle
 \frac{ \mathsf{max\text{-}rb}({i},k-1)
 - \mathsf{last\text{-}rb}({i},k-1)
 }{\mathsf{max\text{-}rb}({i},k-1)}
 &\text{if $\varphi_{i}$ has been played before}
 \\
 0 
 &\text{otherwise}
 \end{cases}
\end{aligned}
\end{equation*}
Here
\begin{math}
 \mathsf{max\text{-}rb}({i},k-1)\Defeq \max\{\mathsf{rb}_{l}\mid l\in[1,k-1], i_{l}=i\}
\end{math} (i.e.\ the greatest $\mathsf{rb}_{l}$ so far, in those attempts  where $\varphi_{i}$ was played), and 
\begin{math}
 \mathsf{last\text{-}rb}({i},k-1)\Defeq
   \mathsf{rb}_{l_{\mathrm{last}}}
\end{math} with
 $l_{\mathrm{last}}$ being the greatest $l\in [1,k-1]$ such that $i_{l}=i$ (i.e.\ the last $\mathsf{rb}_{l}$ for $\varphi_{i}$). 
\end{mydefinition}
Since we try to minimize the robustness values $\mathsf{rb}_{l}$ through falsification attempts, we can expect that $\mathsf{rb}_{l}$ for a fixed arm $\varphi_{i}$ decreases over time. (In the case of the hill-climbing algorithm CMA-ES that we use, this is in fact guaranteed). Therefore the value $ \mathsf{max\text{-}rb}({i},k-1)$ in the definition of $\mathcal{R}(\varphi_{i})$ is the first observed robustness value.  The numerator $ \mathsf{max\text{-}rb}({i},k-1)
 - \mathsf{last\text{-}rb}({i},k-1)$ then represents how much robustness we have reduced so far by hill climbing---hence the name ``hill-climbing gain.'' The denominator $\mathsf{max\text{-}rb}({i},k-1)$ is there for  normalization.

In Algorithm~\ref{algo:main1}, the value $\mathsf{rb}_{k}$ is given by the robustness $ \sem{\M(\bu_{k}), \BoxOp{I}\varphi_{i_{k}}}$. Therefore the MAB choice in Line~\ref{line:MABConj} essentially picks $i_{k}$ for which hill climbing yields greater effect (but  also taking exploration into account---see~\S{}\ref{subsec:MAB}).

In Line~\ref{line:HillClimbConj} we conduct hill-climbing optimization---see~\S{}\ref{subsec:hillClimbingFalsification}. The function $\textsc{Hill-Climb}$  learns from the previous attempts $\bu_{l_{1}},\dotsc, \bu_{l_{m}}$ regarding the same formula $\varphi_{i_{k}}$, and their resulting robustness values $\mathsf{rb}_{l_{1}},\dotsc,\mathsf{rb}_{l_{m}}$. Then it suggests the next input signal $\bu_{k}$ that is likely to minimize the (unknown) function that underlies the correspondences $\bigl[\,\bu_{l_{j}}\mapsto \mathsf{rb}_{l_{j}}\,\bigr]_{j\in [1,m]}$.
 
Lines~\ref{line:HillClimbConj}--\ref{line:updateRbConj} read as follows: the hill-climbing algorithm suggests a single input $\bu_{k}$, which is then selected or rejected (Line~\ref{line:updateRbConj}) based on the robustness value it yields (Line~\ref{line:defRbConj}). We note that this is a simplified picture: in our implementation that uses CMA-ES (it is an evolutionary algorithm), we maintain a population of some ten particles, and each of them is moved multiple times (our choice is three times) before the best one is chosen as $\bu_{k}$.

\subsection{Our MAB-Guided Algorithm II: Disjunctive Safety Properties}\label{subsec:ourAlgorithm2}
The other main algorithm of ours aims to falsify a \emph{disjunctive} safety property $\varphi\equiv \BoxOp{I}(\varphi_{1}\lor\varphi_{2})$. We believe this problem setting is even more important than the conjunctive case, since it encompasses conditional safety properties (i.e.\ of the form $\BoxOp{I}(\varphi_{1}\rightarrow\varphi_{2})$). See~\S{}\ref{subsec:lifeLongProp} for discussions.

In the disjunctive setting, the challenge is that falsification of $\BoxOp{I}\varphi_{i}$ (with $i \in \{1,2\}$) does \emph{not} necessarily imply falsification of $\BoxOp{I}(\varphi_{1}\lor\varphi_{2})$. This is unlike the conjunctive setting. Therefore we need some adaptation of Algorithm~\ref{algo:main1}, so that the two interleaved sequences of falsification attempts for $\varphi_{1}$ and $\varphi_{2}$ are not totally independent of each other. Our solution consists of \emph{restricting} time instants to those where $\varphi_{2}$ is false, in a falsification attempt for $\varphi_{1}$ (and vice versa), in the way described in Def.~\ref{def:robustnessOfSafetyPropertiesRestricted}. 

Algorithm~\ref{algo:main2} shows our MAB-guided algorithm for falsifying a disjunctive safety property $ \BoxOp{I}(\varphi_{1}\lor\varphi_{2})$. The only visible difference is that Line~\ref{line:defRbConj} in Algorithm~\ref{algo:main1} is replaced with Line~\ref{line:defRbConj}'. The new Line~\ref{line:defRbConj}' measures the quality of the suggested input signal $\bu_{k}$ in the way restricted to the region $\mathcal{S}_{k}$ in which the other formula is already falsified. Lem.~\ref{lem:robustSemRestrictionToSubset} guarantees that, if $\mathsf{rb}_{k}<0$, then indeed the input signal $\bu_{k}$ falsifies the original specification $ \BoxOp{I}(\varphi_{1}\lor\varphi_{2})$.

The assumption that makes Alg.~\ref{algo:main2} sensible is that, although it can be hard to find a time instant at which both $\varphi_{1}$ and $\varphi_{2}$ are false (this is required in falsifying $ \BoxOp{I}(\varphi_{1}\lor\varphi_{2})$), falsifying $\varphi_{1}$ (or $\varphi_{2}$) individually is not hard. Without this assumption, the region $\mathcal{S}_{k}$ in Line~\ref{line:defRbConj}' would be empty most of the time. Our experiments in~\S{}\ref{sec:experiments} demonstrate that this assumption is valid in many problem instances, and that Alg.~\ref{algo:main2} is effective.

\section{Experimental Evaluation}\label{sec:experiments} 
We name \mabUcb and \mabEps the two versions of MAB algorithm using strategies $\varepsilon$-Greedy (see Alg.~\ref{algo:greedy}) and UCB1 (see Alg.~\ref{algo:ucb1}). We compared the proposed approach (both versions \mabUcb and \mabEps) with a state-of-the-art falsification framework, namely \breach~\cite{Donze10}.  
\breach encapsulates several hill-climbing optimization algorithms, including \emph{CMA-ES (covariance matrix adaptation evolution strategy)}~\cite{AugerH05}, 
\emph{SA (simulated annealing)}, \emph{GNM (global Nelder-Mead)}~\cite{LuersonLeRiche2004}, etc. 
According to our experience, CMA-ES outperforms other hill-climbing solvers in \breach, 
so the experiments for both \breach and our approach rely on the CMA-ES solver.

Experiments have been executed using Breach 1.2.13 on an Amazon EC2 c4.large instance, 2.9~GHz Intel Xeon E5-2666, 2 virtual CPU cores, 4~GB RAM.

\paragraph{\bf Benchmarks}
We selected three benchmark models from the literature, each one having different specifications. 
The first one is the \emph{Automatic Transmission} (AT) model~\cite{HoxhaAF14,ARCHCOMP19Falsification}. It has two input signals, $\throttle$$\in$$[0,100]$ and $\brake$$\in$$[0,325]$, and computes the car's $\speed$, engine rotation in rounds per minute $\rpm$, and the automatically selected $\gear$. The specifications concern the relation between the three output signals to check whether the car is subject to some unexpected or unsafe behaviors. The second benchmark is the \emph{Abstract Fuel Control} (AFC) model~\cite{JinDKUB14,ARCHCOMP19Falsification}. It takes two input signals, \emph{pedal angle}$\in$$[8.8, 90]$ and \emph{engine speed}$\in$$[900, 1100]$, and outputs the critical signal \emph{air-fuel ratio} ($\AF$), which influences fuel efficiency and car performance. The value is expected to be close to a reference value $\AFref$; $\muSpec$$\equiv$$\nicefrac{|\AF-\AFref|}{\AFref}$ is the deviation of $\AF$ from $\AFref$. The specifications check whether this property holds under both \emph{normal mode} and \emph{power enrichment mode}. The third benchmark is a model of a \emph{magnetic levitation system with a NARMA-L2 neurocontroller} (NN)~\cite{beale1992neural,ARCHCOMP19Falsification}. It takes one input signal, $\mathit{Ref}$$\in$$[1,3]$, which is the reference for the output signal $\mathit{Pos}$, the position of a magnet suspended above an electromagnet. The specifications say that the position should approach the reference signal in a few seconds when these two are not close.

We built the benchmark set \basicBench, as shown in Table~\ref{table:basedBench} that reports  the name of the model and its specifications (ID and formula).
\begin{table}[!tb]
\caption{Benchmark sets \basicBench and \scaledBench}
\label{table:benchmarks}
\centering
\setlength\tabcolsep{2pt}
\begin{subtable}{0.75\textwidth}
\caption{\basicBench (here $\delta_{t'}(\bw)$ represents $\bw^t(t')-\bw^t(0)$).}
\label{table:basedBench}
\resizebox{\textwidth}{!}{
\begin{tabular}{llll}
\toprule
Bench & \multicolumn{2}{c}{Specification} & Parameter\\
\cline{2-3}
 & ID & Formula & \\
\midrule
\multirow{7}{*}{AT} & AT1      & $\BoxOp{[0,30]}((\gear = 3) \rightarrow (\speed > \rho))$              &  $\rho \in \{20.6, 20.4,20.2, 20, 19.8\}$\\
& AT2      &  $\BoxOp{[0,30]}((\gear = 4) \rightarrow (\speed > \rho))$              & $\rho\in\{43, 41, 39, 37, 35\}$          \\
& AT3      &  $\BoxOp{[0,30]}((\gear = 4) \to (\rpm>\rho))$             &   $\rho\in\{700, 800, 900, 1000, 1100\}$        \\
& AT4      & $\BoxOp{[0,30-\tau]}((\delta_{10}(\rpm)>2000) \rightarrow (\delta_\tau(\gear)>0))$              &      $\tau\in\{15, 16, 17, 18, 19\}$             \\
& AT5      & $\BoxOp{[0,30]}((\speed<\rho) \wedge (RPM < 4780))$             &   $\rho \in \{130, 131, 132, 133, 134, 135, 136, 137\}$                \\
& AT6      & $\BoxOp{[0,26]}((\delta_4(\speed)>\rho) \rightarrow (\delta_4(\gear)>0))$ &   $\rho\in\{20, 25, 30, 35, 40\}$                \\
& AT7      &   $\BoxOp{[0,30-\tau]}((\delta_\tau(\speed)>30) \rightarrow (\delta_\tau(\gear)>0))$  &     $\tau\in\{2, 3, 4, 5, 6, 7, 8\}$              \\
\hline
\multirow{2}{*}{AFC}  & AFC1     &   $\BoxOp{[11,50]}((\contrMode = 0) \rightarrow (\muSpec<\rho))$ & $\rho\in\{0.16, 0.17, 0.18, 0.19, 0.2\}$ \\
& AFC2     & $\BoxOp{[11,50]}((\contrMode = 1) \rightarrow (\muSpec<\rho))$   &      $\rho\in\{0.222, 0.224, 0.226, 0.228, 0.23\}$             \\
\hline
%\multirow{3}{*}{NN}                                                & NN1      &\multirow{2}{*}{
%$\begin{array}{l}\BoxOp{[0, 18]}(\neg close\_ref \to reach\_ref\_in\_tau) \\close\_ref\Defeq |Pos - Ref| <= \rho_2+ \rho_1*|Ref|\\  reach\_ref\_in\_tau\Defeq \DiaOp{[0, 2]}(\BoxOp{[0, 1]} (close\_ref)) \end{array}$}
%$\begin{array}{l}\BoxOp{[0, 18]}((\neg (|\mathit{Pos}[t] - \mathit{Ref}[t]| \le p + 0.04*|\mathit{Ref}[t]|))\rightarrow \\ \DiaOp{[0, 2]} (\BoxOp{[0, 1]} (|\mathit{Pos}[t] - \mathit{Ref}[t]| \le p + 0.04*|\mathit{Ref}[t]|)))\end{array}$           
%& $\rho_1 = 0.04, \rho_2\in\{0.001, 0.002, 0.003, 0.004, 0.005\}$ \\
%&       &  
%$\begin{array}{l}\BoxOp{[0, 18]}(\neg close\_ref \to reach\_ref\_in\_tau) \\close\_ref\Defeq |Pos - Ref| <= p+ 0.03*|Ref|\\  reach\_ref\_in\_tau\Defeq \DiaOp{[0, 2]}(\BoxOp{[0, 1]} (close\_ref)) \end{array}$
%$\begin{array}{l}\BoxOp{[0, 18]}((\neg (|\mathit{Pos}[t] - \mathit{Ref}[t]| \le p + 0.03*|\mathit{Ref}[t]|))\rightarrow \\ \DiaOp{[0, 2]} (\BoxOp{[0, 1]} (|\mathit{Pos}[t] - \mathit{Ref}[t]| \le p+ 0.03*|\mathit{Ref}[t]|)))\end{array}$    
%&                  \\ & NN2 &&$\rho_1= 0.03, \rho_2\in\{0.001, 0.002, 0.003, 0.004, 0.005\}$\\
%\hline
 & &\multicolumn{2}{l}{$close\equiv |Pos - Ref| <= \rho+ \alpha*|Ref|$}\\
 & &\multicolumn{2}{l}{$reach\equiv \DiaOp{[0, 2]}(\BoxOp{[0, 1]} (close))$}\\
\multirow{2}{*}{NN}                                                & NN1      & $\BoxOp{[0, 18]}(\neg close \to reach)$, $\alpha = 0.04$ & $\rho\in\{0.001, 0.002, 0.003, 0.004, 0.005\}$ \\
& NN1      & $\BoxOp{[0, 18]}(\neg close \to reach)$, $\alpha = 0.03$ & $\rho\in\{0.001, 0.002, 0.003, 0.004, 0.005\}$\\
\bottomrule
\end{tabular}}
\end{subtable}~~
\begin{subtable}{0.22\textwidth}
\caption{\scaledBench}
\label{table:scaledBench}
\centering
\scriptsize
\resizebox{\textwidth}{!}{
%\begin{tabular}{lcc}
%\toprule
%Original specification ID & scaled output & scaling factors $10^k$\\
%\midrule
%AT1$_i$ with $i \in \{1, \ldots, 5\}$ & $\speed$ & $k\in\{-2, 0, 1, 3\}$\\%$\{10^{-2}, 1, 10, 10^{3}\}$\\
%\hline
%AT5$_i$ with $i \in \{4, \ldots, 8\}$ & $\speed$ & $k\in\{-2, 0, 1, 3\}$\\%$\{10^{-2}, 1, 10, 10^{3}\}$\\
%\hline
%AFC1$_i$ with $i \in \{1, \ldots, 5\}$ & $\muSpec$ & $k\in\{0, 1, 2, 3\}$\\%$\{1, 10, 10^{2}, 10^{3}\}$\\
%\bottomrule
%\end{tabular}}
\begin{tabular}{ccc}
\toprule
Spec ID & scaled           & factor $10^k$                     \\
&  output          &                     \\ \hline
AT1$_1$  & \multirow{5}{*}{$\speed$} & \multirow{5}{*}{$k\in$\{-2,0,1,3\}} \\
AT1$_2$  &                        &                                                  \\
AT1$_3$  &                        &                                                  \\
AT1$_4$  &                        &                                                  \\
AT1$_5$  &                        &                                                  \\ \hline
AT5$_4$  & \multirow{5}{*}{$\speed$} & \multirow{5}{*}{$k\in$\{-2,0,1,3\}} \\
AT5$_5$  &                        &                                                  \\
AT5$_6$  &                        &                                                  \\
AT5$_7$  &                        &                                                  \\
AT5$_8$  &                        &                                                  \\ \hline
AFC1$_1$ & \multirow{5}{*}{$\muSpec$} & \multirow{5}{*}{$k\in$\{0,1,2,3\}}  \\
AFC1$_2$ &                        &                                                  \\
AFC1$_3$ &                        &                                                  \\
AFC1$_4$ &                        &                                                  \\
AFC1$_5$ &                        &                                                 \\
\bottomrule
\end{tabular}}
\end{subtable}
\end{table}
In total, we found 11 specifications. In order to increase the benchmark set and obtain specifications of different complexity, we artificially modified a constant (turned into a parameter named $\tau$ if it is contained in a time interval, named $\rho$ otherwise) of the specification: for each specification $S$, we generated $m$ different versions, named as $S_i$ with $i \in \{1, \ldots, m\}$; the complexity of the specification (in terms of difficulty to falsify it) increases with increasing $i$.\footnote{Note that we performed this classification based on the falsification results of \breach.} In total, we produced 60 specifications. Column {\it parameter} in the table shows which concrete values we used for the parameters $\rho$ and $\tau$. Note that all the specifications but one are disjunctive safety properties (i.e., $\BoxOp{I}(\varphi_{1}\lor\varphi_{2})$), as they are the most difficult case and they are the main target of our approach; we just add AT5 as example of conjunctive safety property (i.e., $\BoxOp{I}(\varphi_{1}\land\varphi_{2})$).

Our approach has been proposed with the aim of tackling the scale problem. Therefore, to better show how our approach mitigates this problem, we generated a second benchmark set \scaledBench as follows. We selected 15 specifications from \basicBench (with concrete values for the parameters) and, for each specification $S$, we changed the corresponding Simulink model by multiplying one of its outputs by a factor $10^k$, with $k \in \{-2,0,1,2,3\}$ (note that we also include the original one using scale factor $10^0$); the specification has been modified accordingly, by multiplying with the scale factor the constants that are compared with the scaled output. We name a specification $S$ scaled with factor $10^k$ as $S^k$. Table~\ref{table:scaledBench} reports the IDs of the original specifications, the output that has been scaled, and the used scaled factors; in total, the benchmark set \scaledBench contains 60 specifications .

\noindent {\bf Experiment} In our context, an {\it experiment} consists in the execution of an approach $A$ (either \breach, \mabEps, or \mabUcb) over a specification $S$ for 30 {\it trials}, using different initial seeds. For each experiment, we record the {\it success} \fr as the number of trials in which a falsifying input was found, and average execution {\it time} of the trials. Complete experimental results are reported in Appendix~A\footnote{The code, models, and specifications are available online at \url{https://github.com/ERATOMMSD/FalStar-MAB}.}. We report aggregated results in Table~\ref{table:aggrResults}.
\begin{table}[!tb]
\caption{Aggregated results for benchmark sets \basicBench and \scaledBench (\fr: \# successes out 30 trials. Time in secs. $\Delta$: percentage difference w.r.t. \breach). Outperformance cases are highlighted, indicated by positive $\Delta$ of SR, and negative $\Delta$ of time.}
\label{table:aggrResults}
\setlength\tabcolsep{1.5pt}
%percentage difference
\resizebox{\textwidth}{!}{
\begin{tabular}{l|rrr|rrr|rrrr|rrrr|rrrr|rrrr}
\toprule
Spec. & \multicolumn{6}{c}{{\tt Breach}} & \multicolumn{8}{c}{\mabEps} & \multicolumn{8}{c}{\mabUcb}\\
ID &  \multicolumn{3}{c}{\fr (/30)}
 & \multicolumn{3}{c}{time (sec.)} & \multicolumn{4}{c}{\fr  (/30)} & \multicolumn{4}{c}{time (sec.)} & \multicolumn{4}{c}{\fr (/30)} & \multicolumn{4}{c}{time (sec.)}\\
& Min & Max & Avg & Min & Max & Avg & Min & Max & Avg & $\Delta$ & Min & Max & Avg & $\Delta$ & Min & Max & Avg & $\Delta$ & Min & Max & Avg & $\Delta$\\
\midrule
AT1 & 14 & 25 & 20.2 & 125 & 361.2 & 223.1 & 24 & 30 & 28.6 & \cellcolor{lightgray} 35.7 & 62.7 & 213.4 & 106.4 & \cellcolor{lightgray}$-$73.4 & 28 & 30 & 29.2 & \cellcolor{lightgray}37.8 & 45.1 & 146.8 & 77.4 & \cellcolor{lightgray}$-$97.1\\
AT2 & 11 & 30 & 20.2 & 14 & 390.6 & 209.8 & 30 & 30 & 30 & \cellcolor{lightgray}43.9 & 11.9 & 126.3 & 54.5 &\cellcolor{lightgray} $-$96.9 & 27 & 30 & 29.4 &\cellcolor{lightgray} 42.2 & 17.7 & 92.5 & 36.8 &\cellcolor{lightgray} $-$112.1\\
AT3 & 29 & 30 & 29.4 & 2.3 & 22.2 & 14.2 & 30 & 30 & 30 & \cellcolor{lightgray}2 & 2.5 & 7 & 3.5 & \cellcolor{lightgray}$-$82.9 & 30 & 30 & 30 & \cellcolor{lightgray}2 & 2.5 & 3.6 & 3 &\cellcolor{lightgray} $-$88.6\\
AT4 & 18 & 30 & 25.8 & 19.5 & 265.3 & 109.6 & 29 & 30 & 29.8 & \cellcolor{lightgray}16 & 7.8 & 45.1 & 24.4 &\cellcolor{lightgray} $-$105 & 30 & 30 & 30 &\cellcolor{lightgray} 16.6 & 6.2 & 36.2 & 22.2 & \cellcolor{lightgray}$-$113.5\\
AT5 & 6 & 23 & 14.1 & 203.1 & 525.9 & 366.2 & 26 & 30 & 28.5 & \cellcolor{lightgray}72.1 & 35.2 & 149 & 93.7 &\cellcolor{lightgray} $-$120.6 & 26 & 30 & 28.2 & \cellcolor{lightgray}71.4 & 37.7 & 154.1 & 99.2 & \cellcolor{lightgray}$-$116.8\\
AT6 & 5 & 29 & 22.8 & 30.1 & 509.5 & 157 & 21 & 30 & 27 &\cellcolor{lightgray} 28 & 2.3 & 300 & 95.1 & \cellcolor{lightgray}$-$98.3 & 22 & 30 & 27 & \cellcolor{lightgray}27.7 & 2.9 & 247.3 & 86.1 &\cellcolor{lightgray} $-$99.4\\
AT7 & 15 & 30 & 26.6 & 12.2 & 314 & 81.5 & 20 & 30 & 28.6 & \cellcolor{lightgray}8.4 & 2.9 & 283.9 & 49.9 &\cellcolor{lightgray} $-$92 & 23 & 30 & 29 &\cellcolor{lightgray} 10.3 & 5.5 & 223.3 & 42.9 & \cellcolor{lightgray}$-$88.3\\
AFC1 & 6 & 30 & 14.4 & 124.8 & 565.6 & 413.5 & 4 & 28 & 12 & $-$28.4 & 171 & 568.4 & 446 & 10.8 & 5 & 30 & 16.4 &\cellcolor{lightgray} 9.7 & 98.7 & 559.8 & 389.9 &\cellcolor{lightgray} $-$9.3\\
AFC2 & 2 & 30 & 18 & 80.7 & 582.3 & 343.4 & 5 & 30 & 20 & \cellcolor{lightgray}23.8 & 43.2 & 547.8 & 301.9 &\cellcolor{lightgray} $-$23.8 & 5 & 30 & 20 &\cellcolor{lightgray} 22.9 & 59.4 & 568.4 & 320.5 & \cellcolor{lightgray}$-$11.1\\
NN1 & 17 & 25 & 20.8 & 212.9 & 384.7 & 292.9 & 14 & 27 & 20.2 & $-$4.5 & 189.5 & 422.8 & 320.3 & 6.2 & 17 & 28 & 22.6 & \cellcolor{lightgray}7.3 & 148.2 & 403.3 & 272.3 & \cellcolor{lightgray}$-$11.8\\
NN2 & 27 & 28 & 27.2 & 55.5 & 93.4 & 73.1 & 30 & 30 & 30 & \cellcolor{lightgray}9.8 & 11 & 39.3 & 26.3 &\cellcolor{lightgray} $-$97.8 & 30 & 30 & 30 & \cellcolor{lightgray}9.8 & 14.6 & 38.2 & 27.4 & \cellcolor{lightgray}$-$92.3\\
\midrule
AT1$^{-2}$ & 30 & 30 & 30 & 42.5 & 97.4 & 56.9 & 28 & 30 & 29 & $-$3.4 & 75.6 & 178.3 & 118.7 & 68.7 & 28 & 30 & 29.4 & $-$2.1 & 54.3 & 136.3 & 80.3 & 33.3\\
AT1$^0$ & 14 & 25 & 20.2 & 125 & 361.2 & 223.1 & 24 & 30 & 28.6 &\cellcolor{lightgray} 35.7 & 62.7 & 213.4 & 106.4 &\cellcolor{lightgray} $-$73.4 & 28 & 30 & 29.2 & \cellcolor{lightgray}37.8 & 45.1 & 146.8 & 77.4 & \cellcolor{lightgray}$-$97.1\\
AT1$^1$ & 4 & 21 & 15.4 & 204.5 & 527.6 & 310.2 & 25 & 30 & 29 &\cellcolor{lightgray} 68.4 & 49 & 234.7 & 102.1 & \cellcolor{lightgray}$-$108 & 27 & 29 & 28.2 & \cellcolor{lightgray}64.5 & 77.5 & 128.7 & 105.1 &\cellcolor{lightgray} $-$93\\
AT1$^3$ & 8 & 24 & 19.8 & 164 & 471.7 & 240.1 & 29 & 30 & 29.8 &\cellcolor{lightgray} 44.6 & 67.5 & 170.6 & 101.9 &\cellcolor{lightgray} $-$77.3 & 29 & 30 & 29.4 & \cellcolor{lightgray}43.4 & 55.4 & 104.8 & 80.6 &\cellcolor{lightgray} $-$93.6\\
AT5$^{-2}$ & 29 & 30 & 29.6 & 61.1 & 163.7 & 102 & 25 & 30 & 27.8 & $-$6.4 & 76.9 & 139.5 & 111.9 & 12.6 & 28 & 30 & 29.4 & $-$0.7 & 48.5 & 131.9 & 85.7 & \cellcolor{lightgray}$-$17\\
AT5$^0$ & 6 & 18 & 11.2 & 291.1 & 525.9 & 423.1 & 28 & 30 & 28.4 & \cellcolor{lightgray}90.5 & 80.2 & 151.3 & 107.4 &\cellcolor{lightgray} $-$117.7 & 26 & 30 & 28 &\cellcolor{lightgray} 89.4 & 68.3 & 154.1 & 114.9 & \cellcolor{lightgray}$-$114.5\\
AT5$^1$ & 0 & 2 & 0.4 & 566.4 & 600 & 593.3 & 27 & 30 & 28.4 & \cellcolor{lightgray}194.8 & 70.7 & 184.5 & 110.3 & \cellcolor{lightgray}$-$138.5 & 25 & 30 & 27.6 & \cellcolor{lightgray}194.1 & 83.1 & 150 & 123.7 &\cellcolor{lightgray} $-$131.2\\
AT5$^3$ & 0 & 1 & 0.2 & 586.4 & 600 & 597.3 & 27 & 30 & 28.6 &\cellcolor{lightgray} 197.2 & 66.8 & 163.3 & 102.5 &\cellcolor{lightgray} $-$142.3 & 27 & 29 & 28 &\cellcolor{lightgray} 197.2 & 80.4 & 160.9 & 111.9 & \cellcolor{lightgray}$-$137.4\\
AFC1$^0$ & 6 & 30 & 14.4 & 124.8 & 565.6 & 413.5 & 4 & 29 & 16.4 &\cellcolor{lightgray} 8.5 & 115.1 & 559.9 & 411.1 &\cellcolor{lightgray} $-$2.8 & 5 & 30 & 16.4 & \cellcolor{lightgray}9.7 & 98.7 & 559.8 & 389.9 & \cellcolor{lightgray}$-$9.3\\
AFC1$^1$ & 7 & 30 & 16.6 & 99 & 548.2 & 393.3 & 3 & 29 & 10.8 & $-$60.9 & 198.1 & 587.6 & 465.8 & 24.6 & 7 & 29 & 17.8 & \cellcolor{lightgray}10.3 & 105.7 & 527.3 & 354.3 &\cellcolor{lightgray} $-$10.3\\
AFC1$^2$ & 0 & 12 & 5.2 & 434.4 & 600 & 535.8 & 3 & 28 & 11.6 &\cellcolor{lightgray} 96.2 & 180.8 & 577.6 & 463 &\cellcolor{lightgray} $-$20.7 & 4 & 30 & 17 & \cellcolor{lightgray}127 & 73.7 & 556.3 & 374.5 &\cellcolor{lightgray} $-$47.3\\
AFC1$^3$ & 1 & 12 & 4.8 & 425.7 & 587.4 & 532.6 & 3 & 30 & 14.4 & \cellcolor{lightgray}109 & 138 & 585.5 & 436.5 & \cellcolor{lightgray}$-$28 & 7 & 30 & 15 &\cellcolor{lightgray} 113 & 77.1 & 553.4 & 403.7 &\cellcolor{lightgray} $-$39.9\\
\bottomrule
\end{tabular}}
\end{table}
For benchmark set \basicBench, it reports aggregated results for each group of specifications obtained from $S$ (i.e., all the different versions $S_i$ obtained by changing the value of the parameter); for benchmark set \scaledBench, instead, results are aggregated for each scaled specification $S^k$ (considering the versions $S^k_i$ obtained by changing the parameter value). We report minimum, maximum and average number of successes \fr, and time in seconds. For \mabEps and \mabUcb, both for \fr and time, we also report the average percentage difference\footnote{$\Delta$=$\nicefrac{((m-b)*100)}{(0.5*(m+b))}$ where $m$ is the result of {\tt MAB} and $b$ the one of \breach.} ($\Delta$) w.r.t. to the corresponding value of \breach.

\noindent{\bf Comparison}
In the following, we compare two approaches $A_1, A_2$ $\in$ \{\breach, \mabEps, \mabUcb\} by comparing the number of their successes \fr and average execution {\it time} using the non-parametric Wilcoxon signed-rank test with 5\% level of significance\footnote{We checked that the distributions are not normal with the non-parametric Shapiro-Wilk test.}~\cite{Wohlin2012}; the null hypothesis is that there is no difference in applying $A_1$ $A_2$ in terms of the compared measure (\fr or time).

\subsection{Evaluation}\label{sec:evaluation}
We evaluate the proposed approach with some research questions.

\researchquestion{Which is the best MAB algorithm for our purpose?}

In \S{}~\ref{subsec:MAB}, we described that the proposed approach can be executed using two different strategies for choosing the arm in the MAB problem, namely \mabEps and \mabUcb. We here assess which one is better in terms of \fr and time. From the results in Table~\ref{table:aggrResults}, it seems that \mabUcb provides slightly better performance in terms of \fr; this has been confirmed by the Wilcoxon test applied over all the experiments (i.e., on the non-aggregated data reported in Appendix~A ): the null hypothesis that using anyone of the two strategies has no impact on \fr is rejected with $p$-value equal to 0.005089, and the alternative hypothesis that \fr is better is accepted with $p$-value=0.9975; in a similar way, the null hypothesis that there is no difference in terms of time is rejected with $p$-value equal to 3.495e-06, and the alternative hypothesis that is \mabUcb is faster is accepted with $p$-value=1. Therefore, in the following RQs, we compare \breach with only the \mabUcb version of our approach.

\researchquestion{Does the proposed approach effectively solve the scale problem?}

We here assess if our approach is effective in tackling the scale problem. Table~\ref{table:expScaling} reports the complete experimental results over \scaledBench for \breach and \mabUcb; for each specification $S$, all its scaled versions are reported in increasing order of the scaling factor.
\begin{table}[!tb]
\caption{Experimental results -- \scaledBench (\fr: \# successes out of 30 trials. Time in secs)}
\label{table:expScaling}
\scriptsize
\centering
\setlength\tabcolsep{3.5pt}
\begin{tabular}{lrrrr|lrrrr|lrrrr}
\toprule
Spec. & \multicolumn{2}{c}{\breach} & \multicolumn{2}{c}{\mabUcb} & Spec. & \multicolumn{2}{c}{\breach} & \multicolumn{2}{c}{\mabUcb} & Spec. & \multicolumn{2}{c}{\breach} & \multicolumn{2}{c}{\mabUcb}\\
%\cline{2-7}
ID	&	\fr	&	time	&	\fr	&	time	&	ID	&	\fr	&	time	&	\fr	&	time	&	ID	&	\fr	&	time	&	\fr	&	time\\
& (/30) & (sec.) & (/30) & (sec.) &&(/30)& (sec.) &(/30)& (sec.) &&(/30)& (sec.) &(/30)&(sec.)\\
\midrule
AT1$_1^{-2}$ & 30 & 51.3 & 30 & 54.3 & AT5$_4^{-2}$ & 30 & 61.1 & 30 & 48.5 & AFC1$_1^0$ & 30 & 124.8 & 30 & 98.7\\
AT1$_1^0$ & 25 & 125 & 29 & 75 & AT5$_4^0$ & 18 & 291.1 & 28 & 94.5 & AFC1$_1^1$ & 30 & 99 & 29 & 105.7\\
AT1$_1^1$ & 20 & 221.1 & 28 & 107.9 & AT5$_4^1$ & 2 & 566.4 & 25 & 150 & AFC1$_1^2$ & 12 & 434.4 & 30 & 73.7\\
AT1$_1^3$ & 23 & 170 & 29 & 55.4 & AT5$_4^3$ & 1 & 586.4 & 28 & 96.2 & AFC1$_1^3$ & 12 & 425.7 & 30 & 77.1\\
\hline
AT1$_2^{-2}$ & 30 & 49 & 29 & 67.5 & AT5$_5^{-2}$ & 30 & 71.3 & 29 & 67.8 & AFC1$_2^0$ & 16 & 421.5 & 23 & 346.8\\
AT1$_2^0$ & 22 & 187.5 & 30 & 45.1 & AT5$_5^0$ & 15 & 369.1 & 27 & 114 & AFC1$_2^1$ & 25 & 345.9 & 27 & 227.9\\
AT1$_2^1$ & 21 & 204.5 & 29 & 77.5 & AT5$_5^1$ & 0 & 600 & 29 & 83.1 & AFC1$_2^2$ & 8 & 497.2 & 25 & 320.5\\
AT1$_2^3$ & 24 & 164 & 30 & 61 & AT5$_5^3$ & 0 & 600 & 27 & 113.8 & AFC1$_2^3$ & 5 & 518.1 & 21 & 364\\
\hline
AT1$_3^{-2}$ & 30 & 42.5 & 30 & 62.4 & AT5$_6^{-2}$ & 29 & 110.2 & 28 & 103.3 & AFC1$_3^0$ & 11 & 457.7 & 15 & 442\\
AT1$_3^0$ & 19 & 239.5 & 29 & 62.5 & AT5$_6^0$ & 10 & 438.2 & 30 & 68.3 & AFC1$_3^1$ & 13 & 479.2 & 14 & 455.5\\
AT1$_3^1$ & 16 & 296.2 & 27 & 128.7 & AT5$_6^1$ & 0 & 600 & 27 & 126.7 & AFC1$_3^2$ & 2 & 590.7 & 15 & 453.2\\
AT1$_3^3$ & 21 & 209.8 & 30 & 93.4 & AT5$_6^3$ & 0 & 600 & 29 & 80.4 & AFC1$_3^3$ & 5 & 545.6 & 8 & 510.6\\
\hline
AT1$_4^{-2}$ & 30 & 44.5 & 30 & 80.8 & AT5$_7^{-2}$ & 30 & 103.6 & 30 & 77.3 & AFC1$_4^0$ & 9 & 498.2 & 9 & 502.1\\
AT1$_4^0$ & 21 & 202.2 & 30 & 57.4 & AT5$_7^0$ & 7 & 491.4 & 26 & 154.1 & AFC1$_4^1$ & 8 & 494 & 12 & 455\\
AT1$_4^1$ & 16 & 301.7 & 28 & 119.5 & AT5$_7^1$ & 0 & 600 & 27 & 134.3 & AFC1$_4^2$ & 4 & 556.8 & 11 & 468.7\\
AT1$_4^3$ & 23 & 185.1 & 29 & 88.3 & AT5$_7^3$ & 0 & 600 & 29 & 108 & AFC1$_4^3$ & 1 & 587.4 & 9 & 513.4\\
\hline
AT1$_5^{-2}$ & 30 & 97.4 & 28 & 136.3 & AT5$_8^{-2}$ & 29 & 163.7 & 30 & 131.9 & AFC1$_5^0$ & 6 & 565.6 & 5 & 559.8\\
AT1$_5^0$ & 14 & 361.2 & 28 & 146.8 & AT5$_8^0$ & 6 & 525.9 & 29 & 143.6 & AFC1$_5^1$ & 7 & 548.2 & 7 & 527.3\\
AT1$_5^1$ & 4 & 527.6 & 29 & 91.9 & AT5$_8^1$ & 0 & 600 & 30 & 124.2 & AFC1$_5^2$ & 0 & 600 & 4 & 556.3\\
AT1$_5^3$ & 8 & 471.7 & 29 & 104.8 & AT5$_8^3$ & 0 & 600 & 27 & 160.9 & AFC1$_5^3$ & 1 & 586 & 7 & 553.4\\
\bottomrule
\end{tabular}
\end{table}
We observe that changing the scaling factor affects (sometimes greatly) the number of successes \fr of \breach; for example, for AT5$_5$ and AT5$_7$ it goes from 30 to 0. For \mabUcb, instead, \fr is similar across the scaled versions of each specification: this shows that the approach is robust w.r.t. to the scale problem as the ``hill-climbing gain'' reward in Def.~\ref{def:hillClimbGain} eliminates the impact of scaling and UCB1 algorithm balances the exploration and exploitation of two sub-formulas. The observation is confirmed by the Wilcoxon test over \fr: the null hypothesis is rejected with $p$-value=1.808e-09, and the alternative hypothesis accepted with $p$-value=1. Instead, the null hypothesis that there is no difference in terms of time cannot be rejected with $p$-value=0.3294. 

\researchquestion{How does the proposed process behave with not scaled benchmarks?}

In RQ2, we checked whether the proposed approach is able to tackle the scale problem for which it has been designed. Here, instead, we are interested in investigating how it behaves on specifications that have not been artificially scaled (i.e., those in \basicBench). From Table~\ref{table:aggrResults} (upper part), we observe that \mabUcb is always better than \breach both in terms of \fr and time, which is shown by the highlighted cases. This is confirmed by Wilcoxon test over \fr and time: null hypotheses are rejected with $p$-values equal to, respectively, 6.02e-08 and 1.41e-08, and the alternative hypotheses that \mabUcb is better are both accepted with $p$-value=1. This means that the proposed approach can also handle specifications that do not suffer from the scale problem, and so it can be used with any kind of specification.

\researchquestion{Is the proposed approach more effective than an approach based on rescaling?}

A na{\"i}ve solution to the scale problem could be to rescale the signals used in specification at the same scale. Thanks to the results of RQ2, we can compare to this possible baseline approach, using the scaled benchmark set \scaledBench. For example, AT5 suffers from the scale problem as $\speed$ is one order of magnitude less than $\rpm$. However, from Table~\ref{table:aggrResults}, we observe that the scaling that would be done by the baseline approach (i.e., running \breach over AT5$^1$) is not effective, as \fr is 0.4/30, that is much lower than the original \fr 14.1/30 of the unscaled approach using \breach. Our approach, instead, raises \fr to 28.4/30 and to 27.6/30 using the two proposed versions. 

The detailed reason is as follows. For example AT5$^1_6$, the specification is 
$$\BoxOp{[0,30]}{(\speed<1350 \land \rpm < 4780)}$$

After artificial scaling of the speed unit ($\times$ 10), the scalings for $\speed$ and $\rpm$ are comparable. Therefore, the baseline approach will consist of simply running \breach for the specification AT5$^1_6$. By monitoring \breach execution, we notice that the na{\"i}ve approach fails because it tries to falsify $\rpm$$<$$4780$, which, however, is not falsifiable; our approach, instead, understands that it must try to falsify $\speed$$<$$\rho$. 

As a result,  our MAB-based falsification for Boolean connectives improves efficiency, even in absence of the scale problem. It does so by taking balance between exploration and exploitation, while original \breach (interpreting $\land$ by infimum) is purely exploiting.

\section{Conclusion and Future work}\label{sec:conclusion}

In this paper, we propose a solution to the \emph{scale problem} that affects falsification of specifications containing Boolean connectives. The approach combines multi-armed bandit algorithms with hill climbing-guided falsification. Experiments show that the approach is robust under the change of scales, and it outperforms a state-of-the-art falsification tool. The approach currently handles binary specifications. As future work, we plan to generalize it to complex specifications having more than two Boolean connectives.

\bibliographystyle{abbrv}
\bibliography{biblio}

\clearpage
\appendix
\section{Experimental results}\label{sec:appendixRes}
%percentage difference
\begin{table}[!h]
\vspace{-45pt}
\caption{Experimental results -- \basicBench (\fr: \# successes out 30 trials. Time in secs. $\Delta$: percentage difference w.r.t. \breach)}
\scriptsize
\centering
\begin{tabular}{lrr|rrrr|rrrr}
\toprule
\multirow{2}{*}{Spec} & \multicolumn{2}{c}{Breach} & \multicolumn{4}{c}{\mabEps} & \multicolumn{4}{c}{\mabUcb} \\
\cline{2-11}
& \fr         & time  & \fr        & time & $\Delta$\fr &  $\Delta$time & \fr        & time  & $\Delta$\fr &  $\Delta$time\\
\midrule
AT1$_1$ & 25 & 125 & 29 & 77.4 & 14.8 & $-$47 & 29 & 75 & 14.8 & $-$50\\
AT1$_2$ & 22 & 187.5 & 30 & 62.7 & 30.8 & $-$99.7 & 30 & 45.1 & 30.8 & $-$122.4\\
AT1$_3$ & 19 & 239.5 & 30 & 91.5 & 44.9 & $-$89.4 & 29 & 62.5 & 41.7 & $-$117.2\\
AT1$_4$ & 21 & 202.2 & 30 & 87.1 & 35.3 & $-$79.5 & 30 & 57.4 & 35.3 & $-$111.6\\
AT1$_5$ & 14 & 361.2 & 24 & 213.4 & 52.6 & $-$51.4 & 28 & 146.8 & 66.7 & $-$84.4\\
\hline
AT2$_1$ & 30 & 14 & 30 & 11.9 & 0 & $-$16.2 & 30 & 17.7 & 0 & 23.3\\
AT2$_2$ & 26 & 96.2 & 30 & 37.8 & 14.3 & $-$87.1 & 30 & 19 & 14.3 & $-$134\\
AT2$_3$ & 20 & 216.3 & 30 & 41.1 & 40 & $-$136.1 & 30 & 23 & 40 & $-$161.5\\
AT2$_4$ & 14 & 331.9 & 30 & 55.2 & 72.7 & $-$143 & 30 & 31.9 & 72.7 & $-$164.9\\
AT2$_5$ & 11 & 390.6 & 30 & 126.3 & 92.7 & $-$102.3 & 27 & 92.5 & 84.2 & $-$123.4\\
\hline
AT3$_1$ & 29 & 21.9 & 30 & 7 & 3.4 & $-$103.6 & 30 & 3.6 & 3.4 & $-$143.6\\
AT3$_2$ & 30 & 2.3 & 30 & 2.5 & 0 & 7.1 & 30 & 2.5 & 0 & 8.3\\
AT3$_3$ & 29 & 22.2 & 30 & 2.6 & 3.4 & $-$158.4 & 30 & 3.4 & 3.4 & $-$146.9\\
AT3$_4$ & 30 & 3 & 30 & 2.8 & 0 & $-$5.8 & 30 & 2.9 & 0 & $-$3.4\\
AT3$_5$ & 29 & 21.7 & 30 & 2.8 & 3.4 & $-$154 & 30 & 2.6 & 3.4 & $-$157.3\\
\hline
AT4$_1$ & 30 & 19.5 & 30 & 7.8 & 0 & $-$85.8 & 30 & 6.2 & 0 & $-$103.5\\
AT4$_2$ & 29 & 48.1 & 30 & 19.5 & 3.4 & $-$84.4 & 30 & 13.3 & 3.4 & $-$113.4\\
AT4$_3$ & 29 & 54.4 & 30 & 31.7 & 3.4 & $-$52.8 & 30 & 29.2 & 3.4 & $-$60.3\\
AT4$_4$ & 23 & 160.5 & 30 & 17.8 & 26.4 & $-$160 & 30 & 36.2 & 26.4 & $-$126.4\\
AT4$_5$ & 18 & 265.3 & 29 & 45.1 & 46.8 & $-$141.9 & 30 & 26.3 & 50 & $-$163.9\\
\hline
AT5$_1$ & 23 & 203.1 & 30 & 35.2 & 26.4 & $-$140.9 & 30 & 37.7 & 26.4 & $-$137.4\\
AT5$_2$ & 16 & 320.4 & 26 & 126.8 & 47.6 & $-$86.6 & 30 & 39.7 & 60.9 & $-$155.9\\
AT5$_3$ & 18 & 290.6 & 30 & 46.6 & 50 & $-$144.7 & 26 & 141.6 & 36.4 & $-$68.9\\
AT5$_4$ & 18 & 291.1 & 27 & 108.5 & 40 & $-$91.4 & 28 & 94.5 & 43.5 & $-$102\\
AT5$_5$ & 15 & 369.1 & 29 & 71.1 & 63.6 & $-$135.4 & 27 & 114 & 57.1 & $-$105.6\\
AT5$_6$ & 10 & 438.2 & 30 & 75.3 & 100 & $-$141.3 & 30 & 68.3 & 100 & $-$146\\
AT5$_7$ & 7 & 491.4 & 28 & 136.8 & 120 & $-$112.9 & 26 & 154.1 & 115.2 & $-$104.5\\
AT5$_8$ & 6 & 525.9 & 28 & 149 & 129.4 & $-$111.7 & 29 & 143.6 & 131.4 & $-$114.2\\
\hline
AT6$_1$ & 28 & 46.4 & 30 & 2.3 & 6.9 & $-$180.8 & 30 & 2.9 & 6.9 & $-$176.5\\
AT6$_2$ & 29 & 30.1 & 30 & 4.3 & 3.4 & $-$149.5 & 30 & 4.3 & 3.4 & $-$150.1\\
AT6$_3$ & 25 & 111.9 & 30 & 9.7 & 18.2 & $-$168.1 & 30 & 11.4 & 18.2 & $-$163\\
AT6$_4$ & 27 & 86.9 & 24 & 159 & $-$11.8 & 58.6 & 23 & 164.8 & $-$16 & 61.8\\
AT6$_5$ & 5 & 509.5 & 21 & 300 & 123.1 & $-$51.8 & 22 & 247.3 & 125.9 & $-$69.3\\
\hline
AT7$_1$ & 30 & 12.2 & 20 & 283.9 & $-$40 & 183.5 & 23 & 223.3 & $-$26.4 & 179.2\\
AT7$_2$ & 15 & 314 & 30 & 33.6 & 66.7 & $-$161.4 & 30 & 43.2 & 66.7 & $-$151.6\\
AT7$_3$ & 25 & 111.9 & 30 & 10.5 & 18.2 & $-$165.6 & 30 & 11.4 & 18.2 & $-$163\\
AT7$_4$ & 28 & 51.1 & 30 & 2.9 & 6.9 & $-$178.6 & 30 & 5.8 & 6.9 & $-$159.5\\
AT7$_5$ & 30 & 13.5 & 30 & 5 & 0 & $-$91.5 & 30 & 5.5 & 0 & $-$84.1\\
AT7$_6$ & 30 & 12.6 & 30 & 5.5 & 0 & $-$78.5 & 30 & 5.6 & 0 & $-$76.2\\
AT7$_7$ & 28 & 54.9 & 30 & 7.6 & 6.9 & $-$151.6 & 30 & 5.6 & 6.9 & $-$162.7\\
\hline
AFC1$_1$ & 30 & 124.8 & 28 & 171 & $-$6.9 & 31.2 & 30 & 98.7 & 0 & $-$23.4\\
AFC1$_2$ & 16 & 421.5 & 15 & 419.2 & $-$6.5 & $-$0.5 & 23 & 346.8 & 35.9 & $-$19.4\\
AFC1$_3$ & 11 & 457.7 & 8 & 506.7 & $-$31.6 & 10.2 & 15 & 442 & 30.8 & $-$3.5\\
AFC1$_4$ & 9 & 498.2 & 5 & 568.4 & $-$57.1 & 13.2 & 9 & 502.1 & 0 & 0.8\\
AFC1$_5$ & 6 & 565.6 & 4 & 564.7 & $-$40 & $-$0.1 & 5 & 559.8 & $-$18.2 & $-$1\\
\hline
AFC2$_1$ & 30 & 80.7 & 30 & 43.2 & 0 & $-$60.5 & 30 & 59.4 & 0 & $-$30.5\\
AFC2$_2$ & 29 & 128.1 & 30 & 100.5 & 3.4 & $-$24.2 & 30 & 123.3 & 3.4 & $-$3.8\\
AFC2$_3$ & 17 & 436.1 & 23 & 326.1 & 30 & $-$28.9 & 24 & 359.3 & 34.1 & $-$19.3\\
AFC2$_4$ & 12 & 489.9 & 12 & 491.9 & 0 & 0.4 & 11 & 492 & $-$8.7 & 0.4\\
AFC2$_5$ & 2 & 582.3 & 5 & 547.8 & 85.7 & $-$6.1 & 5 & 568.4 & 85.7 & $-$2.4\\
\hline
NN1$_1$ & 25 & 221.2 & 27 & 189.5 & 7.7 & $-$15.4 & 28 & 148.2 & 11.3 & $-$39.5\\
NN1$_2$ & 24 & 212.9 & 24 & 212.6 & 0 & $-$0.1 & 28 & 169 & 15.4 & $-$23\\
NN1$_3$ & 19 & 300.1 & 18 & 401.9 & $-$5.4 & 29 & 19 & 308.7 & 0 & 2.8\\
NN1$_4$ & 17 & 384.7 & 18 & 374.6 & 5.7 & $-$2.7 & 21 & 332.2 & 21.1 & $-$14.6\\
NN1$_5$ & 19 & 345.5 & 14 & 422.8 & $-$30.3 & 20.1 & 17 & 403.3 & $-$11.1 & 15.4\\
\hline
NN2$_1$ & 27 & 66.8 & 30 & 11 & 10.5 & $-$143.5 & 30 & 14.6 & 10.5 & $-$128.1\\
NN2$_2$ & 27 & 70.7 & 30 & 17.3 & 10.5 & $-$121.4 & 30 & 23.7 & 10.5 & $-$99.5\\
NN2$_3$ & 28 & 55.5 & 30 & 26 & 6.9 & $-$72.2 & 30 & 27.8 & 6.9 & $-$66.6\\
NN2$_4$ & 27 & 79.1 & 30 & 39.3 & 10.5 & $-$67.3 & 30 & 32.5 & 10.5 & $-$83.5\\
NN2$_5$ & 27 & 93.4 & 30 & 37.8 & 10.5 & $-$84.7 & 30 & 38.2 & 10.5 & $-$83.8\\
\bottomrule
\end{tabular}
\end{table}
\begin{table}[!h]
\caption{Experimental results -- \scaledBench (\fr: \# successes out 30 trials. Time in secs. $\Delta$: percentage difference w.r.t. \breach)}
\scriptsize
\centering
\begin{tabular}{lrr|rrrr|rrrr}
\toprule
\multirow{2}{*}{Spec} & \multicolumn{2}{c}{Breach} & \multicolumn{4}{c}{\mabEps} & \multicolumn{4}{c}{\mabUcb} \\
\cline{2-11}
& \fr         & time  & \fr        & time & $\Delta$\fr &  $\Delta$time & \fr        & time  & $\Delta$\fr &  $\Delta$time\\
\midrule
AT1$_1^{-2}$ & 30 & 51.3 & 30 & 75.6 & 0 & 38.3 & 30 & 54.3 & 0 & 5.6\\
AT1$_1^0$ & 25 & 125 & 29 & 77.4 & 14.8 & $-$47 & 29 & 75 & 14.8 & $-$50\\
AT1$_1^1$ & 20 & 221.1 & 30 & 49 & 40 & $-$127.5 & 28 & 107.9 & 33.3 & $-$68.8\\
AT1$_1^3$ & 23 & 170 & 30 & 82.5 & 26.4 & $-$69.3 & 29 & 55.4 & 23.1 & $-$101.6\\
\hline
AT1$_2^{-2}$ & 30 & 49 & 29 & 115.6 & $-$3.4 & 80.9 & 29 & 67.5 & $-$3.4 & 31.9\\
AT1$_2^0$ & 22 & 187.5 & 30 & 62.7 & 30.8 & $-$99.7 & 30 & 45.1 & 30.8 & $-$122.4\\
AT1$_2^1$ & 21 & 204.5 & 30 & 59.7 & 35.3 & $-$109.6 & 29 & 77.5 & 32 & $-$90.1\\
AT1$_2^3$ & 24 & 164 & 30 & 88.8 & 22.2 & $-$59.5 & 30 & 61 & 22.2 & $-$91.5\\
\hline
AT1$_3^{-2}$ & 30 & 42.5 & 28 & 144.4 & $-$6.9 & 109.1 & 30 & 62.4 & 0 & 38\\
AT1$_3^0$ & 19 & 239.5 & 30 & 91.5 & 44.9 & $-$89.4 & 29 & 62.5 & 41.7 & $-$117.2\\
AT1$_3^1$ & 16 & 296.2 & 30 & 72.3 & 60.9 & $-$121.5 & 27 & 128.7 & 51.2 & $-$78.8\\
AT1$_3^3$ & 21 & 209.8 & 29 & 99.9 & 32 & $-$71 & 30 & 93.4 & 35.3 & $-$76.8\\
\hline
AT1$_4^{-2}$ & 30 & 44.5 & 30 & 79.8 & 0 & 56.8 & 30 & 80.8 & 0 & 57.9\\
AT1$_4^0$ & 21 & 202.2 & 30 & 87.1 & 35.3 & $-$79.5 & 30 & 57.4 & 35.3 & $-$111.6\\
AT1$_4^1$ & 16 & 301.7 & 30 & 94.6 & 60.9 & $-$104.5 & 28 & 119.5 & 54.5 & $-$86.5\\
AT1$_4^3$ & 23 & 185.1 & 30 & 67.5 & 26.4 & $-$93.1 & 29 & 88.3 & 23.1 & $-$70.8\\
\hline
AT1$_5^{-2}$ & 30 & 97.4 & 28 & 178.3 & $-$6.9 & 58.7 & 28 & 136.3 & $-$6.9 & 33.3\\
AT1$_5^0$ & 14 & 361.2 & 24 & 213.4 & 52.6 & $-$51.4 & 28 & 146.8 & 66.7 & $-$84.4\\
AT1$_5^1$ & 4 & 527.6 & 25 & 234.7 & 144.8 & $-$76.8 & 29 & 91.9 & 151.5 & $-$140.7\\
AT1$_5^3$ & 8 & 471.7 & 30 & 170.6 & 115.8 & $-$93.7 & 29 & 104.8 & 113.5 & $-$127.3\\
\hline
AT5$_4^{-2}$ & 30 & 61.1 & 25 & 139.5 & $-$18.2 & 78.1 & 30 & 48.5 & 0 & $-$23\\
AT5$_4^0$ & 18 & 291.1 & 28 & 106.6 & 43.5 & $-$92.8 & 28 & 94.5 & 43.5 & $-$102\\
AT5$_4^1$ & 2 & 566.4 & 29 & 70.7 & 174.2 & $-$155.6 & 25 & 150 & 170.4 & $-$116.2\\
AT5$_4^3$ & 1 & 586.4 & 28 & 89.3 & 186.2 & $-$147.1 & 28 & 96.2 & 186.2 & $-$143.6\\
\hline
AT5$_5^{-2}$ & 30 & 71.3 & 27 & 113.8 & $-$10.5 & 45.9 & 29 & 67.8 & $-$3.4 & $-$5.1\\
AT5$_5^0$ & 15 & 369.1 & 28 & 98.2 & 60.5 & $-$115.9 & 27 & 114 & 57.1 & $-$105.6\\
AT5$_5^1$ & 0 & 600 & 27 & 115.4 & 200 & $-$135.5 & 29 & 83.1 & 200 & $-$151.4\\
AT5$_5^3$ & 0 & 600 & 30 & 66.8 & 200 & $-$159.9 & 27 & 113.8 & 200 & $-$136.2\\
\hline
AT5$_6^{-2}$ & 29 & 110.2 & 29 & 76.9 & 0 & $-$35.5 & 28 & 103.3 & $-$3.5 & $-$6.5\\
AT5$_6^0$ & 10 & 438.2 & 28 & 100.5 & 94.7 & $-$125.4 & 30 & 68.3 & 100 & $-$146\\
AT5$_6^1$ & 0 & 600 & 29 & 90.8 & 200 & $-$147.4 & 27 & 126.7 & 200 & $-$130.3\\
AT5$_6^3$ & 0 & 600 & 30 & 70 & 200 & $-$158.2 & 29 & 80.4 & 200 & $-$152.7\\
\hline
AT5$_7^{-2}$ & 30 & 103.6 & 28 & 116.1 & $-$6.9 & 11.4 & 30 & 77.3 & 0 & $-$29.1\\
AT5$_7^0$ & 7 & 491.4 & 30 & 80.2 & 124.3 & $-$143.9 & 26 & 154.1 & 115.2 & $-$104.5\\
AT5$_7^1$ & 0 & 600 & 30 & 90 & 200 & $-$147.8 & 27 & 134.3 & 200 & $-$126.8\\
AT5$_7^3$ & 0 & 600 & 27 & 123.3 & 200 & $-$131.8 & 29 & 108 & 200 & $-$139\\
\hline
AT5$_8^{-2}$ & 29 & 163.7 & 30 & 113 & 3.4 & $-$36.7 & 30 & 131.9 & 3.4 & $-$21.6\\
AT5$_8^0$ & 6 & 525.9 & 28 & 151.3 & 129.4 & $-$110.6 & 29 & 143.6 & 131.4 & $-$114.2\\
AT5$_8^1$ & 0 & 600 & 27 & 184.5 & 200 & $-$105.9 & 30 & 124.2 & 200 & $-$131.4\\
AT5$_8^3$ & 0 & 600 & 28 & 163.3 & 200 & $-$114.4 & 27 & 160.9 & 200 & $-$115.4\\
\hline
AFC1$_1^0$ & 30 & 124.8 & 29 & 115.1 & $-$3.4 & $-$8.1 & 30 & 98.7 & 0 & $-$23.4\\
AFC1$_1^1$ & 30 & 99 & 29 & 198.1 & $-$3.4 & 66.7 & 29 & 105.7 & $-$3.4 & 6.5\\
AFC1$_1^2$ & 12 & 434.4 & 28 & 180.8 & 80 & $-$82.4 & 30 & 73.7 & 85.7 & $-$142\\
AFC1$_1^3$ & 12 & 425.7 & 30 & 138 & 85.7 & $-$102.1 & 30 & 77.1 & 85.7 & $-$138.7\\
\hline
AFC1$_2^0$ & 16 & 421.5 & 23 & 331.7 & 35.9 & $-$23.8 & 23 & 346.8 & 35.9 & $-$19.4\\
AFC1$_2^1$ & 25 & 345.9 & 12 & 456.8 & $-$70.3 & 27.6 & 27 & 227.9 & 7.7 & $-$41.1\\
AFC1$_2^2$ & 8 & 497.2 & 15 & 446.6 & 60.9 & $-$10.7 & 25 & 320.5 & 103 & $-$43.2\\
AFC1$_2^3$ & 5 & 518.1 & 16 & 438.9 & 104.8 & $-$16.5 & 21 & 364 & 123.1 & $-$34.9\\
\hline
AFC1$_3^0$ & 11 & 457.7 & 16 & 531.3 & 37 & 14.9 & 15 & 442 & 30.8 & $-$3.5\\
AFC1$_3^1$ & 13 & 479.2 & 7 & 514.9 & $-$60 & 7.2 & 14 & 455.5 & 7.4 & $-$5.1\\
AFC1$_3^2$ & 2 & 590.7 & 6 & 554.3 & 100 & $-$6.4 & 15 & 453.2 & 152.9 & $-$26.3\\
AFC1$_3^3$ & 5 & 545.6 & 16 & 472.1 & 104.8 & $-$14.4 & 8 & 510.6 & 46.2 & $-$6.6\\
\hline
AFC1$_4^0$ & 9 & 498.2 & 4 & 559.9 & $-$76.9 & 11.7 & 9 & 502.1 & 0 & 0.8\\
AFC1$_4^1$ & 8 & 494 & 3 & 571.7 & $-$90.9 & 14.6 & 12 & 455 & 40 & $-$8.2\\
AFC1$_4^2$ & 4 & 556.8 & 6 & 555.9 & 40 & $-$0.2 & 11 & 468.7 & 93.3 & $-$17.2\\
AFC1$_4^3$ & 1 & 587.4 & 7 & 547.8 & 150 & $-$7 & 9 & 513.4 & 160 & $-$13.4\\
\hline
AFC1$_5^0$ & 6 & 565.6 & 10 & 517.5 & 50 & $-$8.9 & 5 & 559.8 & $-$18.2 & $-$1\\
AFC1$_5^1$ & 7 & 548.2 & 3 & 587.6 & $-$80 & 6.9 & 7 & 527.3 & 0 & $-$3.9\\
AFC1$_5^2$ & 0 & 600 & 3 & 577.6 & 200 & $-$3.8 & 4 & 556.3 & 200 & $-$7.6\\
AFC1$_5^3$ & 1 & 586 & 3 & 585.5 & 100 & $-$0.1 & 7 & 553.4 & 150 & $-$5.7\\
\bottomrule
\end{tabular}
\end{table}

\end{document}